\documentstyle[preprint,aps,epsfig,equation]{revtex}
\begin{document}
\newif\iftightenlines\tightenlinesfalse
\tightenlines\tightenlinestrue
%
\def\lsim{\:\raisebox{-0.5ex}{$\stackrel{\textstyle<}{\sim}$}\:}
\def\gsim{\:\raisebox{-0.5ex}{$\stackrel{\textstyle>}{\sim}$}\:}
\newcommand{\tanb}{\mbox{$\tan \! \beta$}}
\newcommand{\be}{\begin{equation}}
\newcommand{\ben}{\begin{subequations}}
\newcommand{\een}{\end{subequations}}
\newcommand{\beq}{\begin{eqalignno}}
\newcommand{\eeq}{\end{eqalignno}}
\newcommand{\ee}{\end{equation}}
\renewcommand{\thefootnote}{\fnsymbol{footnote}} 
\def\llgm{\left\lgroup\matrix}
\def\rrgm{\right\rgroup}
\draft
\preprint{\vbox{\hbox{IFT--99/068}\hbox{PM/99-34}\hbox{SLAC-PUB-8249}}}
\title{Supersymmetric Higgs Boson Pair Production: Discovery Prospects
at Hadron Colliders} 
\author{A.\ Belyaev $^{1,2}$,  
        M.\ Drees $^{1,3}$, J.\ K.\ Mizukoshi $^{4}$}
\address{$^1$ Instituto de F\'{\i}sica Te\'orica,
              Universidade  Estadual Paulista, \\
              Rua Pamplona 145, 01405--900, S\~ao Paulo, Brazil. \\
         $^2$ Skobeltsyn Institute for Nuclear Physics,
              Moscow State University, 119 899, Moscow, Russia\\
         $^3$ Lab de Physique Mathematique, Univ. Montpellier II,
              Montpellier, France \\
         $^4$ Stanford Linear Accelerator Center, University of Stanford, \\ 
              Stanford, CA 94309, USA}
\date{\today}
\maketitle

\begin{abstract}
We study the potential of hadron colliders in the search for the pair
production of neutral Higgs bosons in the framework of the Minimal
Supersymmetric Standard Model.  We perform a detailed signal and
background analysis, working out efficient kinematical cuts for the
extraction of the signal. The important role of squark loop
contributions to the signal is re--emphasized.  If the signal is
sufficiently enhanced by these contributions, it could even be
observable at the next run of the upgraded Tevatron collider in the
near future. At the LHC the pair production of light and heavy Higgs
bosons might be detectable simultaneously.
\end{abstract}


\section{Introduction}

The search for Higgs bosons is one of the most important tasks for
experiments at present and future high energy colliders
\cite{higrev}. At $e^+e^-$ colliders, precision measurements of many
of their properties are in principle quite straightforward, given
sufficient energy and luminosity \cite{lcrev}. However, the LEP
collider is nearing the end of its lifetime without having detected
any signal for Higgs bosons. Various plans for the construction of
future (linear) $e^+e^-$ colliders exist, but the prospects for their
realization remain uncertain. On the other hand, the Tevatron will
soon start its next collider run with slightly increased beam energy
and greatly increased luminosity; a few years later experiments at the
LHC will commence taking data. It is therefore imperative to fully
exploit the opportunities for Higgs boson searches at hadron
colliders.

Unfortunately most signals for the production of neutral Higgs bosons
at hadron colliders suffer from a low signal to background ratio,
and/or use decay modes with very low branching ratios. Examples for
the former case are associated $WH$ or $ZH$ production followed by $H
\rightarrow b \bar{b}$ decays \cite{wh}, while an example for the
latter situation is single Higgs boson production followed by $H
\rightarrow Z Z^* \rightarrow 4 \ell$ decays, if $m_H < 2 M_Z$.  The
discovery of Higgs bosons at hadron colliders is therefore not
straightforward; this is certainly true for the Tevatron, but also
holds for the LHC, with the possible exception of one or two
``gold--plated'' modes (e.g., $W H$ \cite{whgg} or $t\bar{t}H$
\cite{gun_tth} production, followed by $H \rightarrow \gamma \gamma$ 
decays, leading to $\ell \nu \gamma \gamma X$ final states)
which however have small cross sections. The study of
their properties at hadron colliders is correspondingly even more
difficult. It is therefore important to search for Higgs bosons in as
many different final states as possible. On the one hand multiple
positive signals will increase confidence that a Higgs boson has
indeed been found, since different final states will have different
systematic uncertainties, a major concern for final states with low
signal to background ratio. Moreover, by comparing the strengths of
(or bounds on) several signals, much can be learned about the
couplings of the Higgs bosons.

In this paper we study the production of two neutral Higgs bosons in
gluon fusion, followed by the decays of both bosons into $b \bar{b}$
pairs. We focus on the final states where both Higgs bosons have
(nearly) the same mass, since the resulting kinematical constraint
helps to reduce the background. The SM cross section for this process
has been computed some time ago \cite{smhh}; it was found to be too
small to be useful. However, the scalar sector of the SM suffers from
well--known naturalness problems. These can be cured by introducing
Supersymmetry.  Here we concentrate on the simplest potentially
realistic supersymmetric model, the minimal supersymmetric standard
model (MSSM) \cite{mssmrev}.  Several effects can greatly enhance the
Higgs pair production cross section in the MSSM as compared to the SM:

\begin{itemize}

\item The MSSM contains two Higgs doublet superfields, and correspondingly
two independent vacuum expectation values (vevs). While the sum of the
square of these vevs is fixed by the well--known masses of the $W$ and $Z$
bosons, their ratio, denoted by \tanb, is a free parameter of the MSSM.
If $\tanb \gg 1$, the Yukawa coupling of the $b-$quark is enhanced by a
factor $\sim \tanb$ compared to its SM value. It thus becomes comparable
to the top quark Yukawa coupling for $\tanb \sim m_t(m_t)/m_b(m_t) \simeq 60$,
which is possible in most realizations of the MSSM. For Higgs boson masses
around 100 GeV the squared $b-$loop contribution then exceeds the $t-$loop
contribution, which is suppressed by the large mass of the top quark, by
a factor $\sim 15$ \cite{plehn}.

\item The MSSM contains three physical neutral Higgs bosons. If CP is
conserved in the scalar sector of the theory, these states can be
classified as two CP--even scalars $h, \ H$ (with $m_h < m_H$) and one
CP--odd pseudoscalar $A$. For some region of parameter space ($m_A \sim
300$ GeV, $\tanb \lsim 4$) the branching ratio for $H \rightarrow hh$
decays is sizable. $h$ pair production through resonant $H$ exchange
is then enhanced by a factor $(g M_W/\lambda_t \Gamma_H)^2 \sim 100$,
where the $Hhh$ coupling is ${\cal O}(g M_W)$, $g$ being the $SU(2)$
gauge coupling, and $\lambda_t$ is the top Yukawa coupling 
\cite{gun_hh,plehn}.

\item In addition to mandating the existence of (at least) two Higgs
doublets, Supersymmetry also requires the presence of superpartners of
all known quarks and leptons. This gives rise to new squark loop
contributions to Higgs boson pair production through gluon fusion.  In
particular, contributions from loops involving $\tilde{b}$ or
$\tilde{t}$ squarks can exceed those from $b$ and $t$ quark loops by
more than two orders of magnitude \cite{bdemn}. This enhancement can
occur for all values of $m_A$ and \tanb, but requires a fairly light
squark mass eigenstate ($\tilde{t}_1$ or $\tilde{b}_1$), as well as
large trilinear Higgs--squark--squark couplings.

\end{itemize}

In this paper we show that neutral Higgs boson pair production can
indeed be discovered in the $4b$ final state at the LHC, if at least
one of these possible enhancement factors is large. Under favorable
circumstances, even the next run of the Tevatron collider could yield
a signal. These results are based on a complete Monte Carlo simulation
of signal and background, including realistic $b-$tagging
efficiencies, parton showering and hadronization, and a simple
parameterization of resolution smearing. The statistically most
significant signal almost always comes from the $hh$ final state, in
some cases augmented by $hA$ and $AA$ final states (for $m_A \simeq
m_h$), but $HH$ production can also give a detectable signal.
The $4b-$jet signature for MSSM Higgs bosons has first been
studied in refs.\cite{gun_hh}, with emphasis on $(h,H,A) b \bar{b}$ as well
as $H \rightarrow hh,AA \rightarrow 4b$ production.  These earlier
references did not study Higgs pair production in the continuum, however.

The remainder of this article is organized as follows. In Sec.~II we
describe our Monte Carlo simulation. Sec.~III deals with the
calculation of signal and background, and discusses the cuts used to
maximize the statistical significance of the signal. These results are
used in Sec.~IV to estimate the discovery reach of the Tevatron and
the LHC, and Sec.~V contains our conclusions.

\section{Monte Carlo Simulation}

In order to study the observability of the signal for Higgs pair
production in the $4b$ final state, we have written MC generators for
complete sets of signal as well as background processes. These generators
were designed as new external user defined processes for the
PYTHIA~5.7/JETSET~7.4 package~\cite{pythia} using the standard PYTHIA
routine PYUPEV.

A total of 6 processes contribute to the signal ($hh, \ hH, \ hA, \
HH, \ HA$ and $AA$ production). FORTRAN codes of the corresponding
squared matrix elements, including squark loop contributions to $hh, \
hH, \ HH$ and $AA$ production, had already been available
\cite{bdemn}.

We used the CompHEP package \cite{comp} to generate background events
on the parton level. This is based on a calculation of exact
tree--level matrix elements for $gg, q \bar{q} \rightarrow b \bar{b} b
\bar{b}$, as well as $gg, q \bar{q} \rightarrow Z b \bar{b}$ followed
by $Z \rightarrow b \bar{b}$ decays.

For both signal and background, the effects of initial and final state
radiation, hadronization, as well as decay of the $b-$flavored hadrons
have been taken into account. For simplicity we used the independent
fragmentation scheme. CTEQ4L parton distributions \cite{cteq} have been used
everywhere. Since our matrix element calculation only includes the
lowest nontrivial order in QCD perturbation theory, there is a
considerable scale dependence. We chose the same value $Q$ for
factorization and renormalization scales, including the scale of the
running Yukawa couplings. For Higgs pair production this scale $Q$ was
set to the average mass of the Higgs bosons in the final state, while
for the $Zb \bar{b}$ background $Q= M_Z$ has been taken. Finally, when
calculating the $4b$ pure QCD background, we took $Q$ to be the
average $b \bar{b}$ pair invariant mass. This choice of a rather high
scale $Q$ should be conservative. While the predictions for both
signal $S$ and background $B$ would be higher for smaller values of
$Q$, leaving the ratio $S/B$ more or less the same, the significance
$S/\sqrt{B}$ increases with decreasing $Q$.

PYTHIA demands that the scales for initial and final state radiation
off user defined processes should be set explicitly by the user.
These scales were chosen equal to factorization and renormalization
scales.

In our analysis we used the cone algorithm for jet reconstruction,
with cone size $\Delta R= \sqrt{\Delta\varphi^2 +
\Delta\eta^2}=0.5$. The minimum $E_T$ threshold for a cell to be
considered as a jet initiator has been chosen as 3~GeV, while the
minimal summed $E_T$ for a collection of cells to be accepted as a jet
has been set at 7~GeV; the cell size $\delta \eta \times \delta \phi$
has been taken as $8/25 \times \pi/12$.  Finally, the energy of each
jet was smeared, with resolution $\Delta E/E=0.8/\sqrt{E}$ ($E$ in GeV)
for the Tevatron and $\Delta E/E=0.5/\sqrt{E}$ ($E$ in GeV) for the
LHC. The mentioned resolutions are in agreement with those used in the
Supersymmetry/Higgs RUN II workshop\cite{d0tag} and ATLAS/CMS studies
of high $p_T \ b-$jets (see, for example, \cite{lhc}), respectively.


Finally, we mention that the use of on--shell quark masses has been
advocated when calculating the Yukawa couplings \cite{plehn}, based on
experience with two--loop corrections to quark loop contributions to
single Higgs production cross sections \cite{nlo1}. This would
increase the signal cross section in the region of large \tanb, where
the dominant contributions involve the bottom Yukawa coupling, by more
than a factor of five. However, our signal cross section is quartic in
Yukawa couplings. If QCD corrections are of similar size as for single
Higgs production, they could be absorbed by choosing a running $b$
mass at a scale intermediate between $m_b$ and $m_H$. Moreover, QCD
corrections to the all--important squark loop contributions are not
even known for the case of single Higgs production, if $m_{\tilde q}
\lsim \sqrt{\hat{s}}/2$. Finally, our background calculation does not
include (possibly quite large) QCD corrections, either. For these
reasons, we prefer to conservatively quote results based on running
Yukawa couplings taken at the large scale $Q$ introduced in the
previous paragraph.


\section{Signal and Background Study}

\subsection{Signal and Background Rates}

As well known, the masses of the CP--even Higgs bosons in the MSSM
receive large radiative corrections, in particular from the top--stop
sector \cite{yanagida}. We treat these corrections in the one--loop
approximation, as calculated in the effective potential method
\cite{ellis}. We include leading two--loop corrections by choosing the
appropriate scales for the running top quark mass when calculating the
masses of the neutral Higgs bosons \cite{dn2}.  As in
ref.\cite{bdemn}, we take equal soft breaking contributions to
diagonal entries of the stop and sbottom mass matrices
($m_{\tilde{t}_L} = m_{\tilde{t}_R} = m_{\tilde{b}_R} \equiv
m_{\tilde{q}}$), as well as equal trilinear soft breaking parameters
in the stop and sbottom sectors ($A_t = A_b \equiv A_q$). We fix the
running masses of the top and bottom quarks to $m_t(m_t) = 165$ GeV
and $m_b(m_b) = 4.2$ GeV, respectively. This leaves us with a total of
5 free parameters which determine our signal cross sections: $m_A, \
\tanb, \ m_{\tilde q},\ A_q$ and the supersymmetric higgsino mass
parameter $\mu$.

Of course, this parameter space is subject to experimental constraints.
We interpret the unsuccessful searches for Higgs bosons at LEP as
implying the following constraints \cite{lephiggs}:
\beq \label{e1n}
m_h &> 95 \ {\rm GeV}, \ \ \ {\rm if}\ |\sin(\alpha-\beta)| > 0.6; 
\nonumber \\
m_h + m_A &> 180 \ {\rm GeV}, \ \ {\rm if}\ |\cos(\alpha-\beta)| > 0.6;
\nonumber \\
m_h & > 80 \ {\rm GeV}.
\eeq
Here, $\alpha$ is the mixing angle in the CP--even Higgs sector;
$\sin(\alpha-\beta)$ and $\cos(\alpha-\beta)$ determine the size of
the $ZZh$ and $ZAh$ couplings, respectively \cite{hunter}. The third
condition in (\ref{e1n}) has been included to catch ``pathological''
cases. These constraints are only a rough approximation of the current
search limits; however, a more elaborate treatment of the experimental
constraints would not change our results significantly.

We also demand that the masses of the lighter physical stop and
sbottom states satisfy
\be \label{e2n}
m_{\tilde{t}_1}, \ m_{\tilde{b}_1} > 90 \ {\rm GeV},
\ee
which follows from squark searches at LEP \cite{lepsquark}. We ignore
constraints from stop and sbottom searches at the Tevatron, since
these are stronger than (\ref{e2n}) only for very large mass splitting
between the squark mass eigenstates and the lightest superparticle
(LSP).  We also require that the contribution from stop and sbottom
loops to the electroweak $\rho-$parameter \cite{delrho} satisfies
\cite{langacker}
\be \label{e3n}
\delta \rho_{\tilde{t} \tilde{b}} \leq 0.0017.
\ee
Finally, we only consider values of $A_q$ and $\mu$ in the range
\be \label{e4n}
|A_q|, \ |\mu| \leq 3 m_{\tilde q};
\ee
this is necessary to avoid the breaking of electric charge and color
in the absolute minimum of the scalar potential \cite{falsevac}.

\begin{figure}[htb]  
\protect
\mbox{\epsfig{file=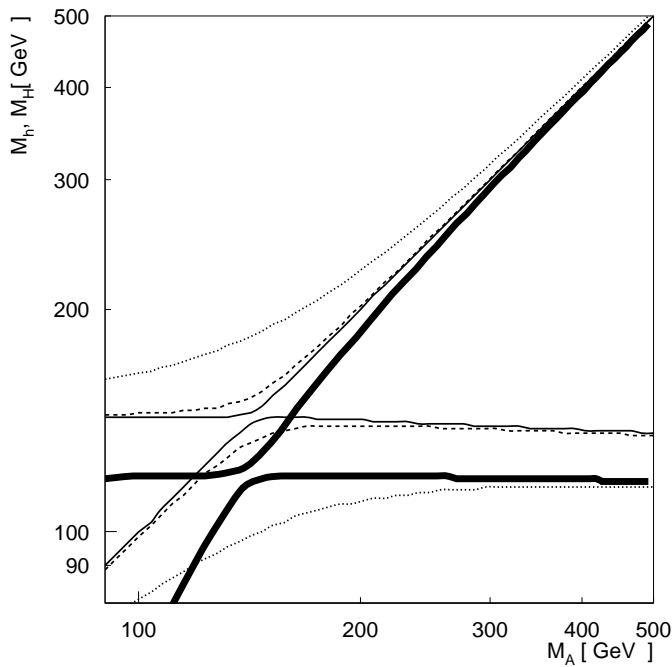,width=0.57\textwidth}}
\caption{$m_h$ and $m_H$ versus $m_A$ for $m_{\tilde q} = 1$ TeV, $A_q
= 2.4$ TeV, $\mu=0$ and $\tan\beta$=2 (dotted), 9 (dashed) and 45
(thin solid), respectively. The heavy solid line is for $m_{\tilde q}
= 330$ GeV, $A_q=700$ GeV, $\mu=600$ GeV and $\tanb=50$. Note that we
did not impose the Higgs search limits (1) here.}
\label{fig:mh-ma}
\end{figure}

As already mentioned, there are 6 different channels for producing two
neutral Higgs bosons in the MSSM: $HH$, $hh$, $AA$, $Hh$, $HA$ and
$hA$. Often several channels contribute to a given signal even after
cuts have been applied, once the experimental resolution has been
taken into account. The reason is that often two Higgs bosons are
essentially degenerate in mass, especially for high \tanb.  For
example, for heavy squarks and $\tanb \gg 1$, one has $m_h \simeq m_A$
for $m_A \lsim 120$ GeV and $m_H \simeq m_A$ for $m_A \gsim 120$
GeV. In the `intermediate' region (120 GeV $\lsim m_A \lsim$ 130 GeV)
all three neutral Higgs bosons are almost degenerate. This is
illustrated in Fig.~\ref{fig:mh-ma}, which shows $(m_h,m_H)$ versus
$m_A$ for three values of \tanb. In Fig.~\ref{fig:mh-ma} we fixed
$m_{\tilde q},\ A_q$ and $\mu$ to 1 TeV, 2.4 TeV and 0, respectively,
corresponding to `maximal' (stop) mixing. On the other hand, even at
large \tanb\ there are cases where this degeneracy is not very close,
if $m_A$ is not large. This is illustrated by the heavy line in
Fig.~\ref{fig:mh-ma}, where we have taken a set of parameters leading
to large $\tilde{b}$ loop contributions to $hh$ production: $m_{\tilde
q} = 330$ GeV, $A_q = 700$ GeV, $\mu = 600$ GeV and $\tanb = 50$. For
this choice of parameters, $m_h$ is significantly smaller than $m_A$
everywhere, and $m_H\simeq m_A$ only for $m_A \gsim 250$ GeV.

\begin{figure}[htb]
\protect
\hspace*{-2cm}
\mbox{\epsfig{file=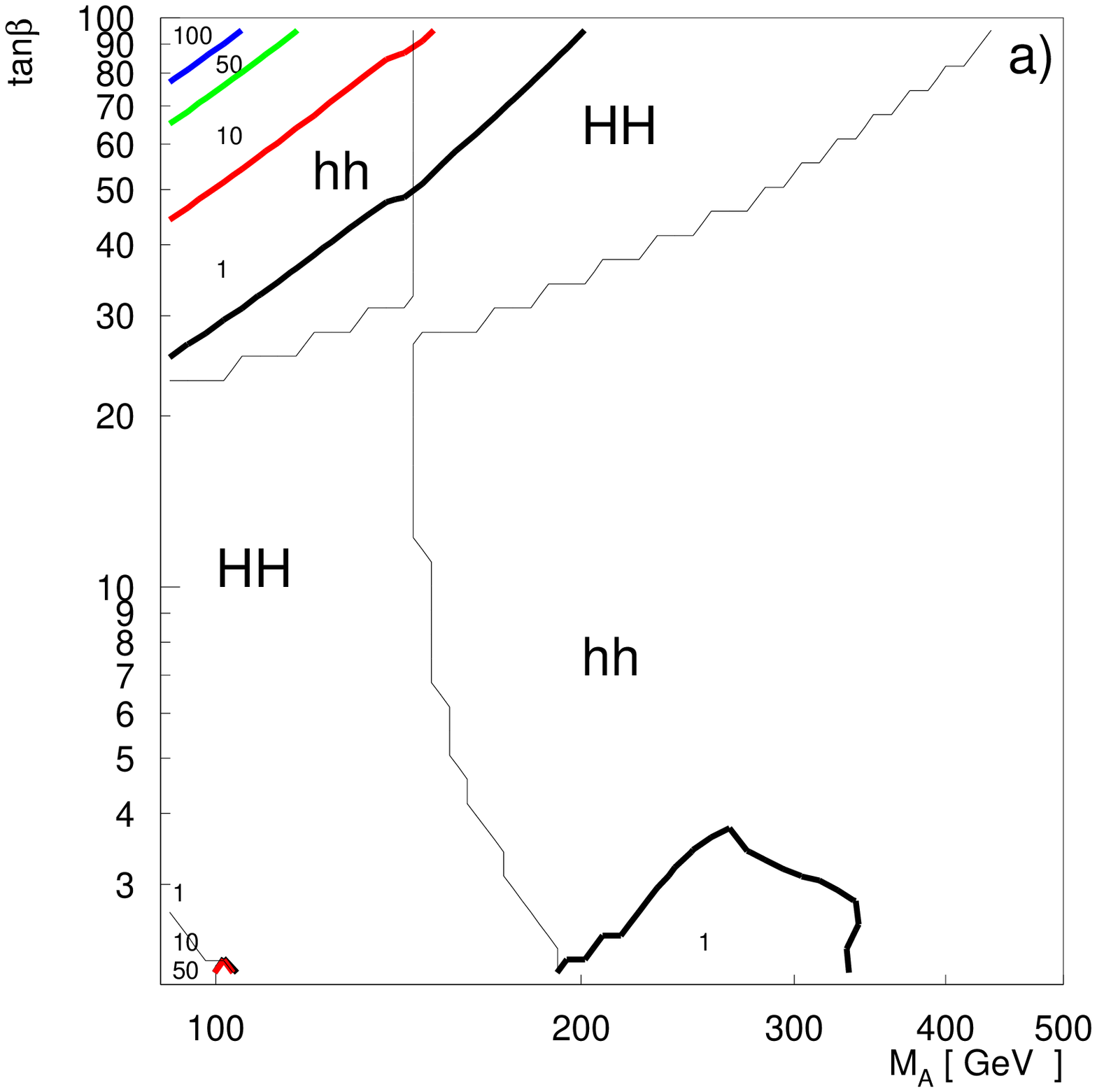,width=0.57\textwidth}
      \epsfig{file=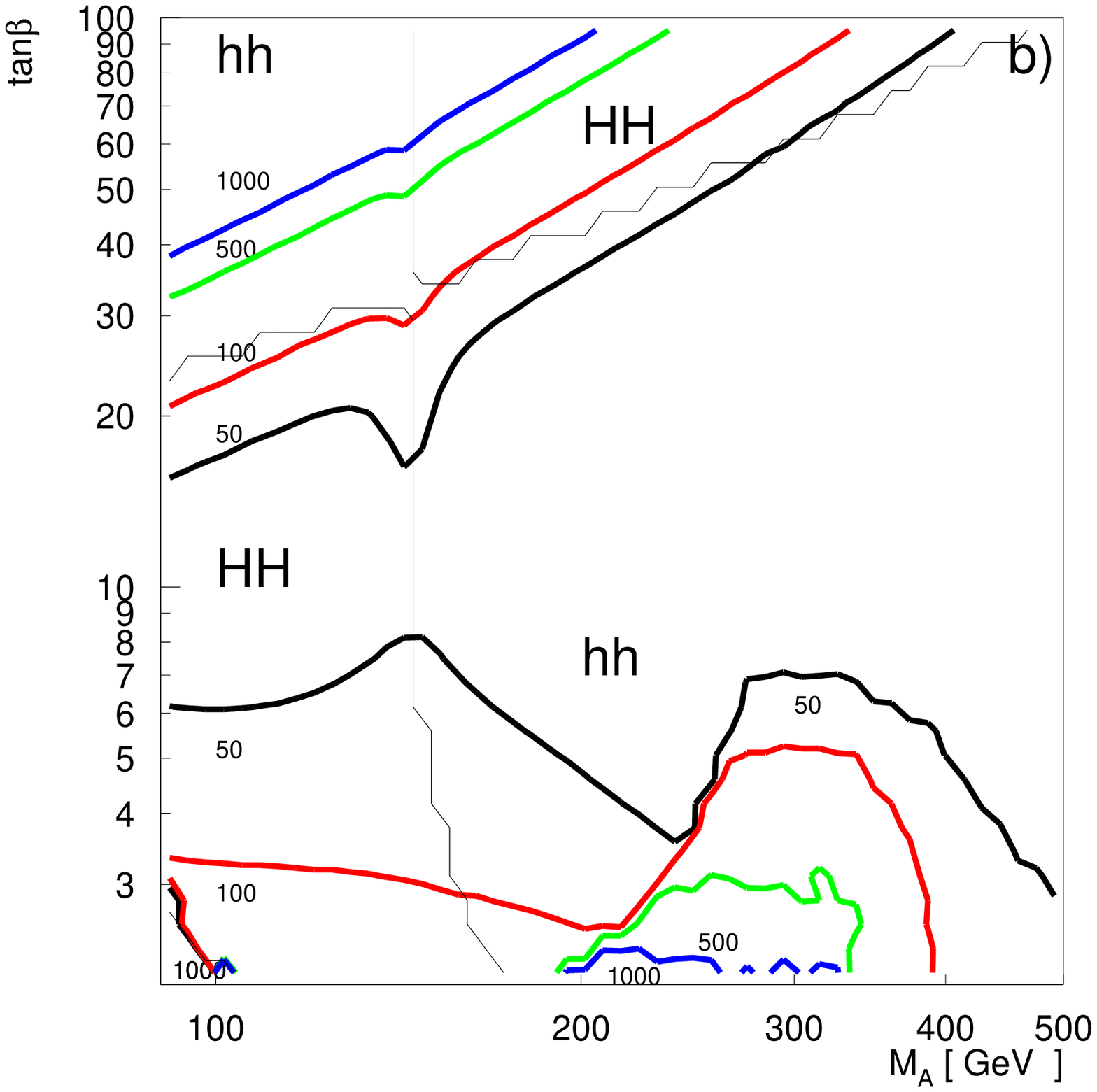,width=0.57\textwidth}}
\\
\vspace*{-7cm}
\\
\hspace*{3cm}Tevatron \ \ \hspace*{8cm}LHC\\
\hspace*{3cm}no squarks\hspace*{8cm}no squarks
\\
\vspace*{5.5cm}
\caption{Contours of constant cross section (in fb) for combined Higgs
pair production channels, for the case of negligible squark loop
contributions. Results for the Tevatron and the LHC are shown in
frames (a) and (b), respectively.}
\label{fig:cslevel}
\end{figure}

In our analysis we have combined contributions from different
production channels assuming a Gaussian distribution for the
reconstructed Higgs boson mass. We start with the diagonal process
($hh, \ HH$ or $AA$ production) giving the best signal significance,
and then add all other contributions to the ``search window'' defined
below, after resolution smearing has been taken into account. In order
to give an idea of the signal rate for negligible squark loop
contributions, in Fig.~\ref{fig:cslevel} we present contours of
constant total signal cross section in fb in the $(m_A,\tan\beta)$
plane.

The squark mass parameters used in Fig.~\ref{fig:cslevel} are the same
as used for the thin lines in Fig.~\ref{fig:mh-ma}. Two of the three
possible enhancement factors mentioned in the Introduction are clearly
visible in this figure: the enhancement of $b-$quark loop
contributions at large \tanb, where the cross section grows $\propto
\tan^4 \beta$; and the enhancement around $m_A = 300$ GeV at small
\tanb, from $H \rightarrow h h$ decays. The total cross section is
about 200 times higher at the LHC than at the Tevatron. Given an
integrated luminosity of 100 fb$^{-1}$, we expect well over 1,000
Higgs pair events at the LHC for all combinations of $m_A$ and
\tanb. In contrast, if squark loop contributions are indeed small, at
the Tevatron the raw signal rate is often too small to give a positive
signal at Run II (2 fb$^{-1}$), or even at TeV33 (25 fb$^{-1}$), even
if backgrounds were negligible.

The thin lines in Fig.~\ref{fig:cslevel} delineate regions where
different channels dominate the total signal. For large $m_A$ and not
very large \tanb, $hh$ production is always the most important
channel. However, for $m_A \gsim 130$ GeV the enhancement of the
$b-$quark Yukawa coupling is only felt by the heavy Higgs bosons $H$
and $A$, which are nearly degenerate at large \tanb; the production of
these heavier states therefore dominates in this region. For smaller
values of $m_A$, $hh$ production is dominant at large \tanb, where the
$h b \bar{b}$ coupling is proportional to \tanb.  However, for $\tanb
\lsim 20$ and small $m_A, \ HH$ production dominates again, since the
$h t \bar{t}$ coupling is suppressed ($\propto \cot \! \beta$).

In order to decide whether a Higgs pair cross section leads to a
detectable signal, we have to compute the background rate.  In order
to suppress ``fake'' backgrounds, we require that all four $b-$jets
are tagged as such. The total cross sections for the two main
irreducible backgrounds, from $Z b \bar{b}$ production as well as pure
QCD $b \bar{b} b \bar{b}$ production, for the basic parton--level
acceptance cuts $p_T>25$ GeV, $\Delta R_{jj}>0.5$ are:
\begin{eqalignno*}
\sigma[Zb \bar{b}] ( Q= M_Z) &= 1.5 \;(59)
\;\; \mbox{pb at the Tevatron (LHC)}; 
\\ 
\sigma[b\bar b b \bar b]( Q=M_{b{\bar b}}) &= 2.6 \;(330) 
\;\; \mbox{pb at the Tevatron (LHC)}.
\end{eqalignno*}
Note that these cross sections do not include any $b-$tagging
efficiency.  For the same acceptance cuts, and again ignoring
$b-$tagging efficiencies, the cross sections for the most important
``fake'' backgrounds from $Wb\bar b$ and especially $jjb\bar{b}$
production are:
\begin{eqalignno*}
\sigma[jjb \bar{b} ]( Q= M_{b{\bar b}}) &= 1.6 \; (164) 
\;\; \mbox{nb at the Tevatron (LHC)}; 
\\
\sigma[Wb\bar{b}] (Q= M_W) &= 3.1\; (19.1) 
\;\; \mbox{pb at the Tevatron (LHC)}.
\end{eqalignno*}
Since the mis--tag probability of light quark and gluon jets is
expected to be $\lsim 1\%$ \cite{d0tag,cmstag}, after $b-$tagging
these ``fake'' backgrounds are much smaller than the irreducible
backgrounds listed above. We therefore ignore these ``fake''
backgrounds from now on.

The background cross sections quoted above only include contributions
where the entire final state is produced in a single partonic
collision. In particular at the LHC, there are also contributions
where two independent partonic collisions can produce one $b \bar{b}$
pair each. The partons in the initial state can come either from the
same $pp$ collision, or from independent collisions; recall that at
full luminosity, one expects about 20 inelastic $pp$ interactions per
bunch collision at the LHC. In the former case, the cross section can
be estimated as $\sigma(gg, q \bar{q} \rightarrow b \bar{b})^2/(2 \pi
R^2)$; here, $\pi R^2 \simeq 15$ mb is the ``effective area'' of the
proton \cite{cdf_mult}. We find that this contribution amounts to no
more than 10\% of the $2 \rightarrow 4$ pure QCD background, and can
thus be ignored. In contrast, we found the $4b$ background from
two independent $pp$ interactions to be comparable to that from QCD $2
\rightarrow 4$ processes. However, good $b-$tagging requires a precise
determination of both the primary vertex (from the partonic collision)
and the secondary vertices (from $b-$decay). Background events where
the two $b \bar{b}$ pairs come from independent $pp$ collisions should
have two distinct ``primary'' vertices. This feature should allow to
suppress these backgrounds efficiently. In the following we therefore
only include irreducible backgrounds from a single partonic collision.

If only the basic acceptance cuts are used, these irreducible
backgrounds are clearly far larger than the signal that can be
achieved without sizable contribution from squark loops, see
Fig.~\ref{fig:cslevel}.  A more elaborate set of cuts is thus
necessary; this is the topic of the following subsection.

\subsection{Kinematical Analysis}
 
The signal consists of two pairs of $b-$jets. As already noted, we
require all four $b-$jets to be tagged as such, in order to suppress
``fake'' backgrounds. A realistic description of the $b-$tagging
efficiency is therefore very important. In case of the Tevatron, we
use the projected $b-$tagging efficiency of the upgraded D\O\ detector
\cite{d0tag}:
\be \label{ed0tag}
\epsilon_b = 0.57 \cdot \tanh \left( \frac {p_T} {35 \ {\rm GeV} }
\right),
\ee
where $p_T$ refers to the transverse momentum of the $b-$jet. For the LHC,
we parameterize numerical results by the CMS collaboration \cite{cmstag}:
\be \label{ecmstag}
\epsilon_b = \left\{ \begin{array} {l l}
0.6, & \ \ \ {\rm for} \ p_T > 100 \ {\rm GeV} \\
0.1 + p_T/(200 \ {\rm GeV}), & \ \ \ {\rm for} \ 40 \ {\rm GeV} \leq p_T
\leq 100 \ {\rm GeV} \\
1.5 p_T/(100 \ {\rm GeV}) - 0.3, & \ \ \ {\rm for} \ 25 \ {\rm GeV} \leq
p_T \leq 40 \ {\rm GeV}
\end{array} \right.
\ee
We assume that $b-$jets can be tagged only for pseudorapidity
$|\eta_b| \leq 2$ by both Tevatron and LHC experiments.

\begin{figure}
\vspace*{-0.5cm}
\noindent
\mbox{\epsfig{file=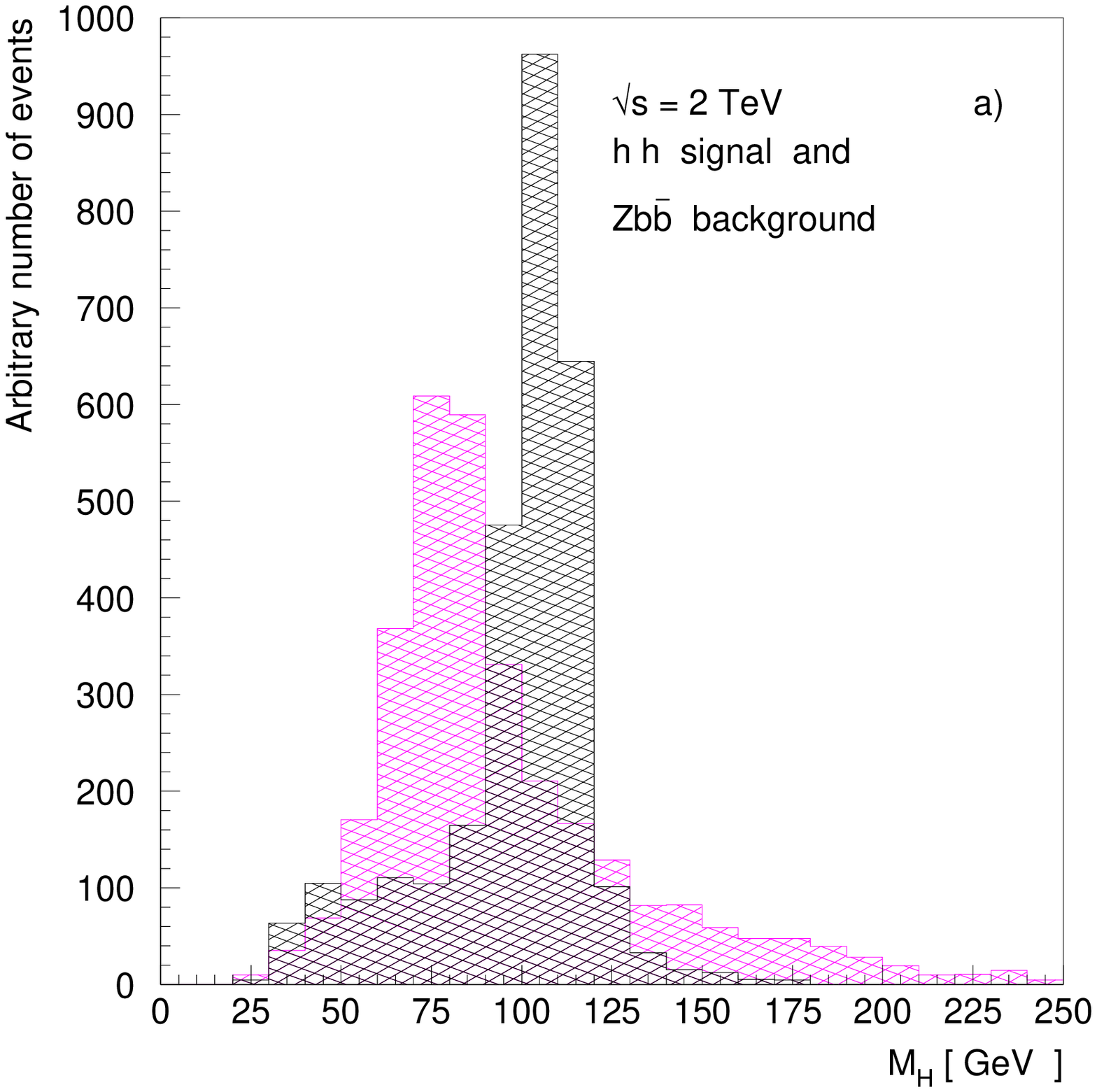 ,width=0.45\textwidth}
      \epsfig{file=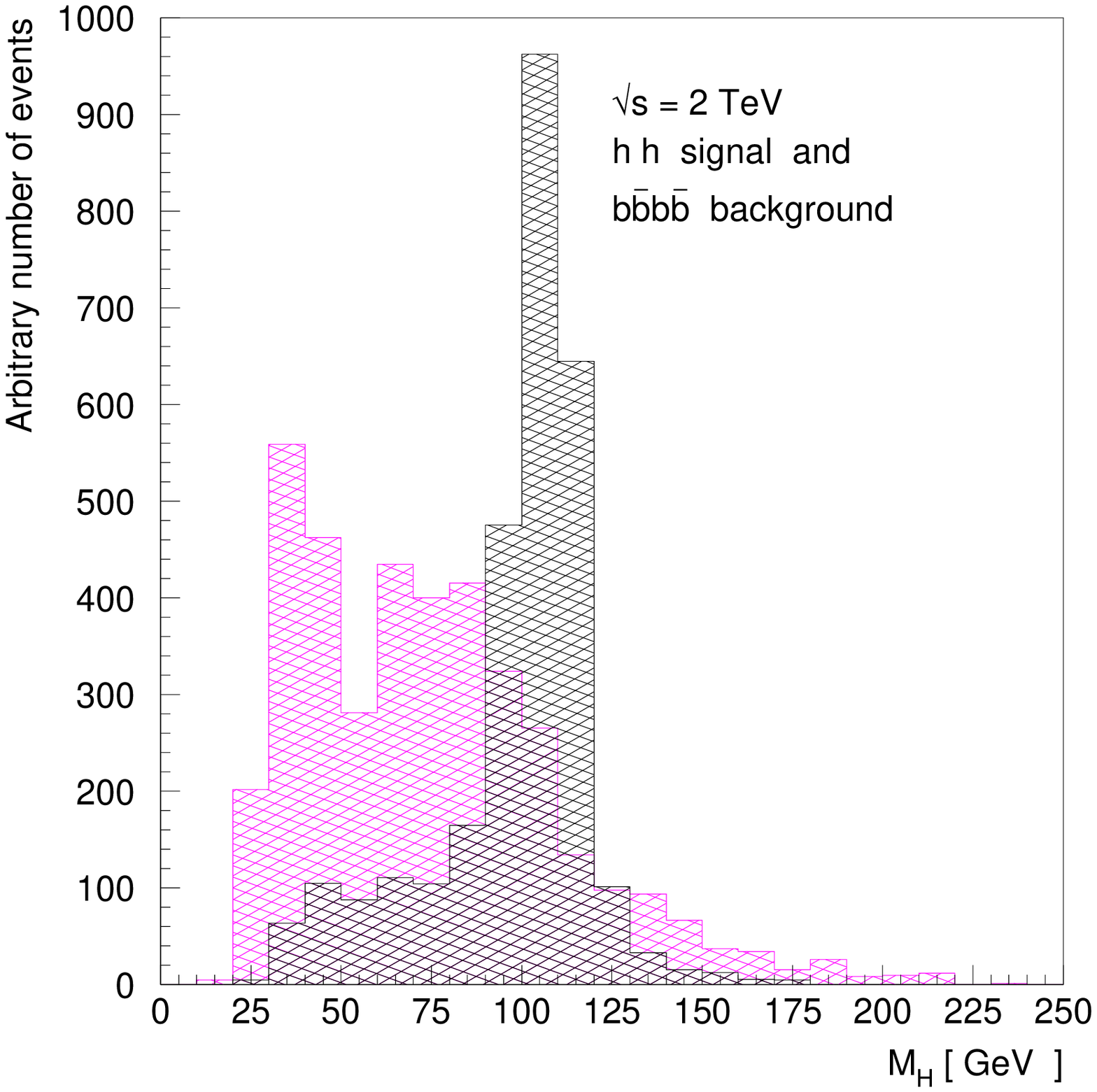,width=0.45\textwidth}}
\vspace*{-0.5cm}
\mbox{\epsfig{file=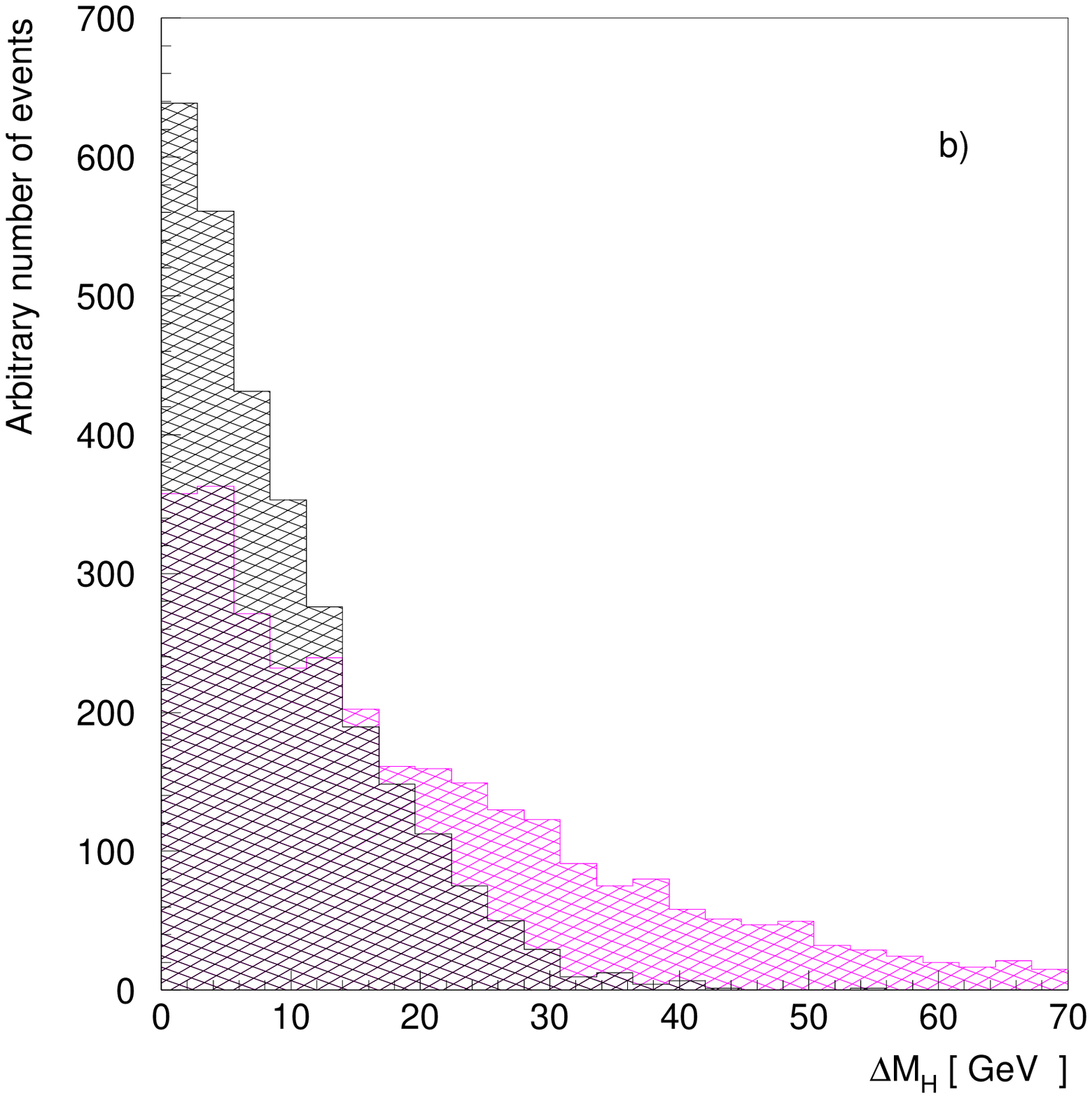 ,width=0.45\textwidth}
      \epsfig{file=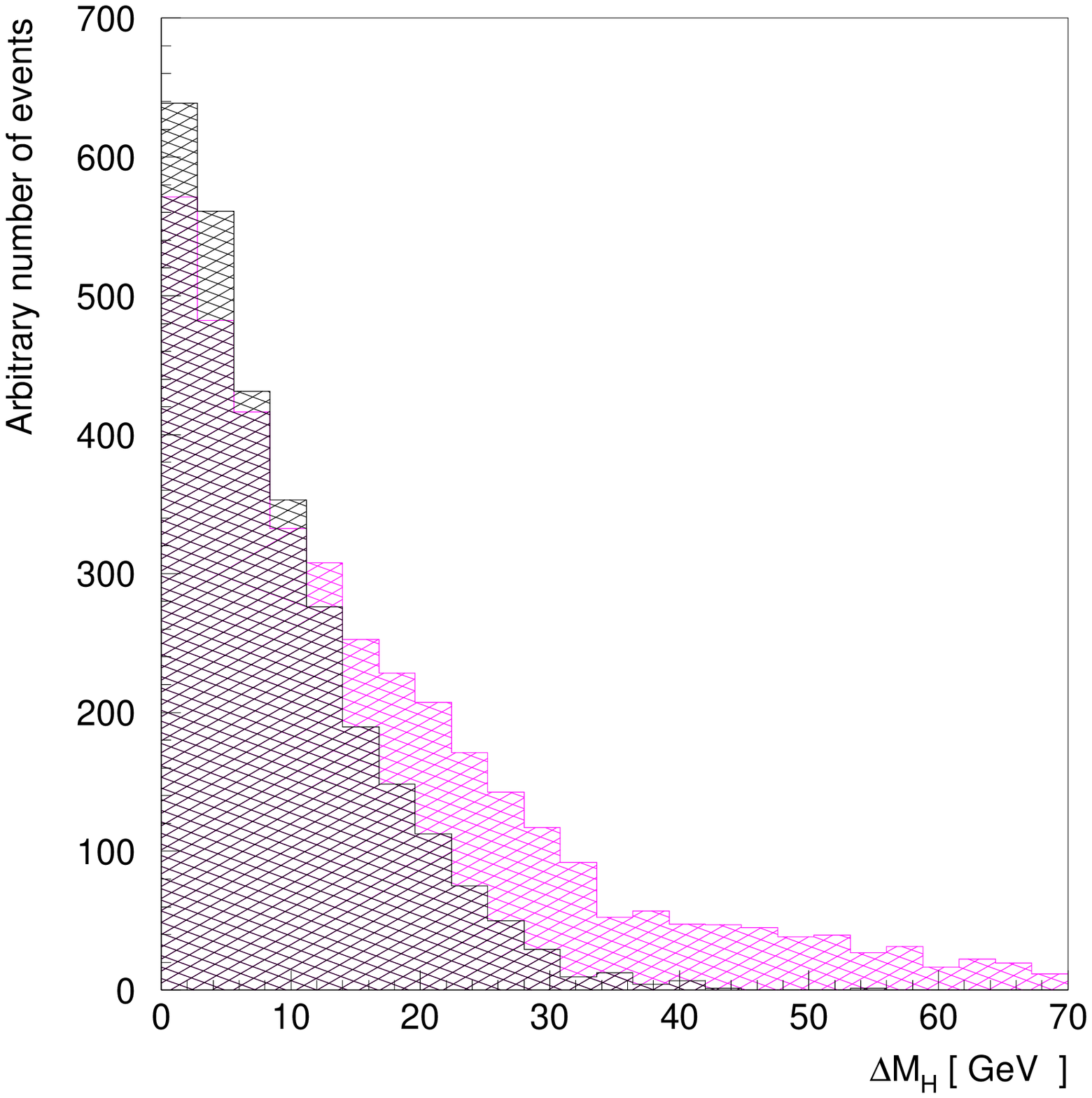,width=0.45\textwidth}}
\vspace*{-0.5cm}
\mbox{\epsfig{file=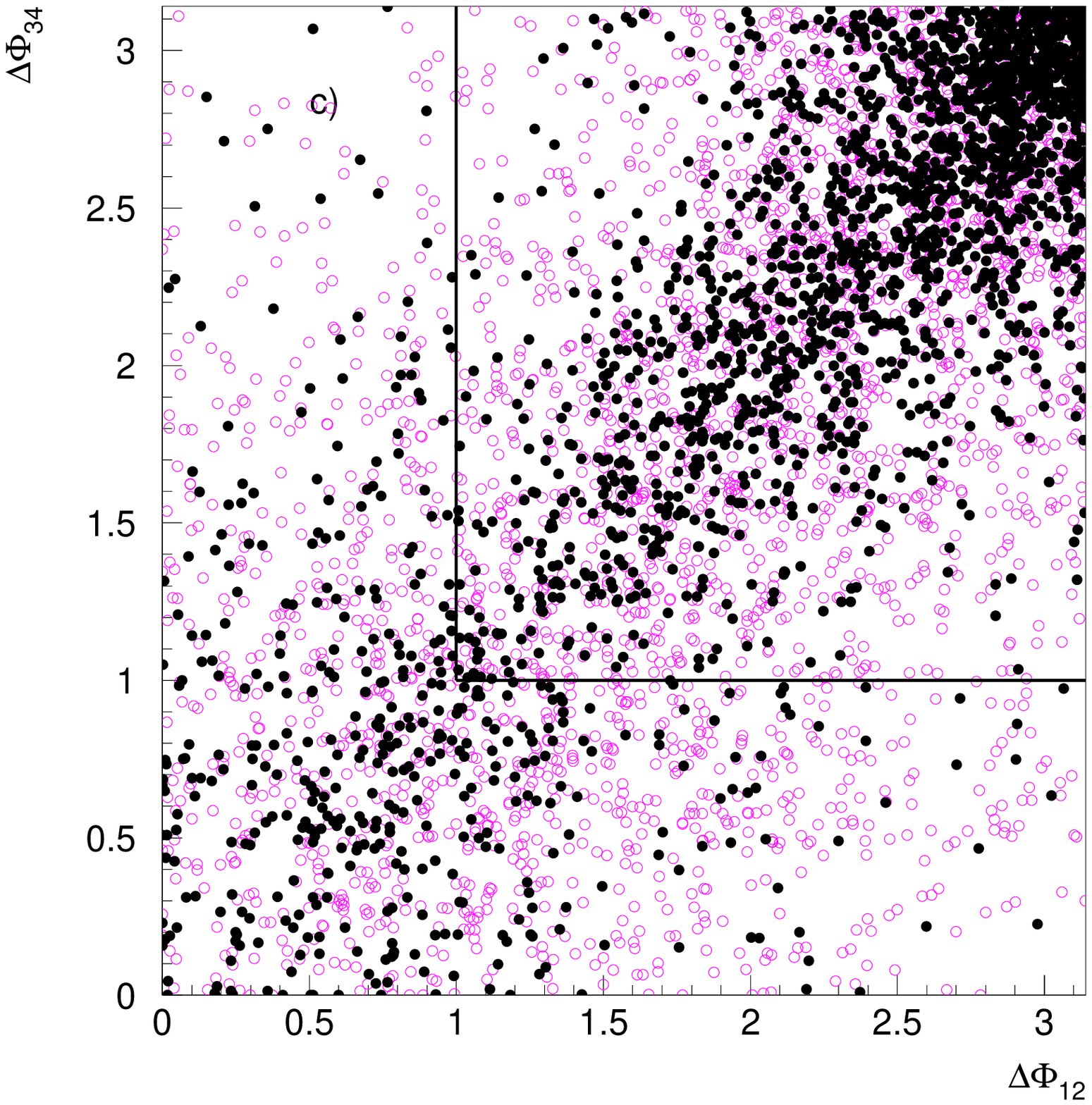 ,width=0.45\textwidth}
      \epsfig{file=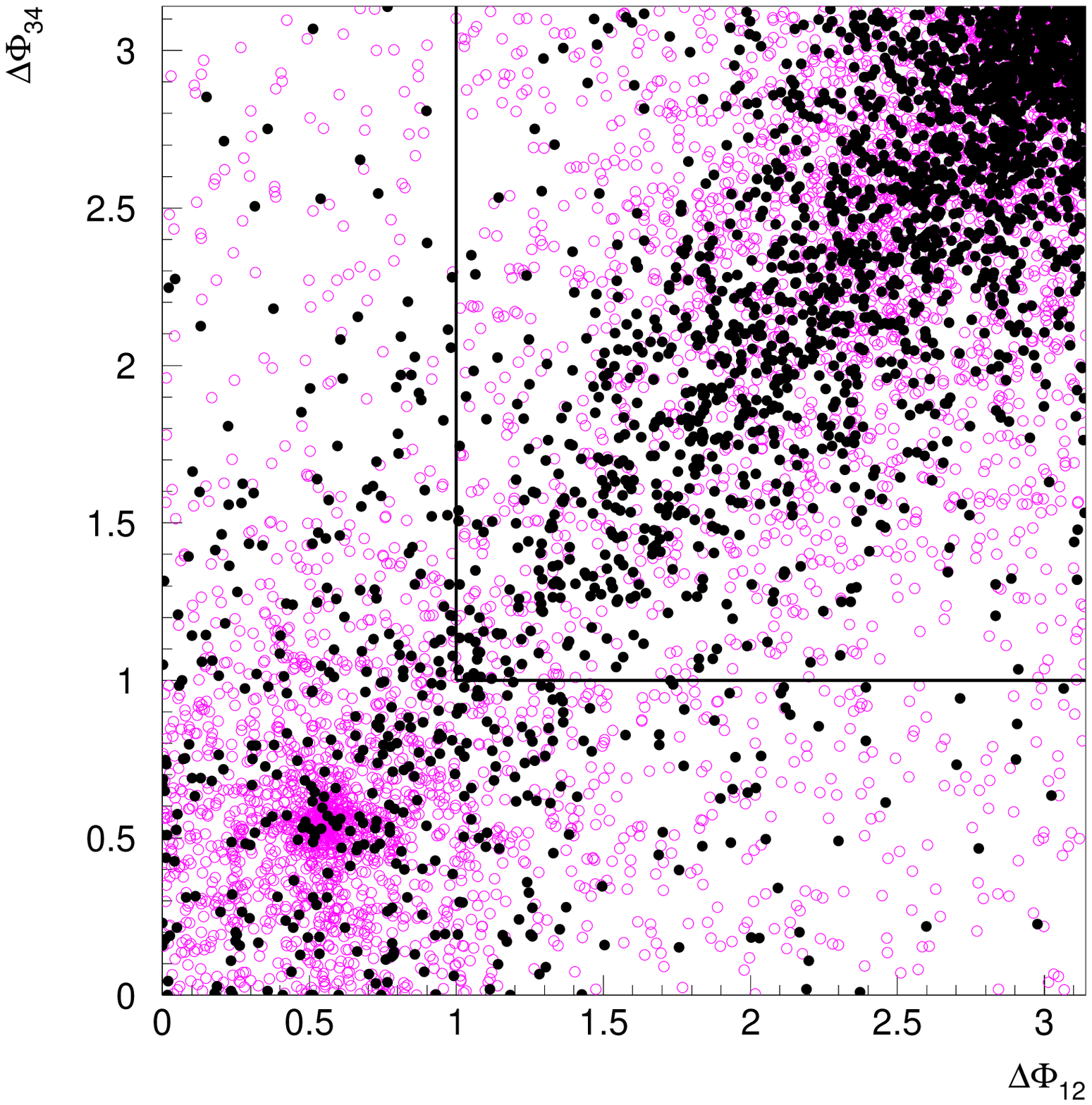,width=0.45\textwidth}}
\newpage
\vspace*{-0.5cm}
\noindent
\mbox{\epsfig{file=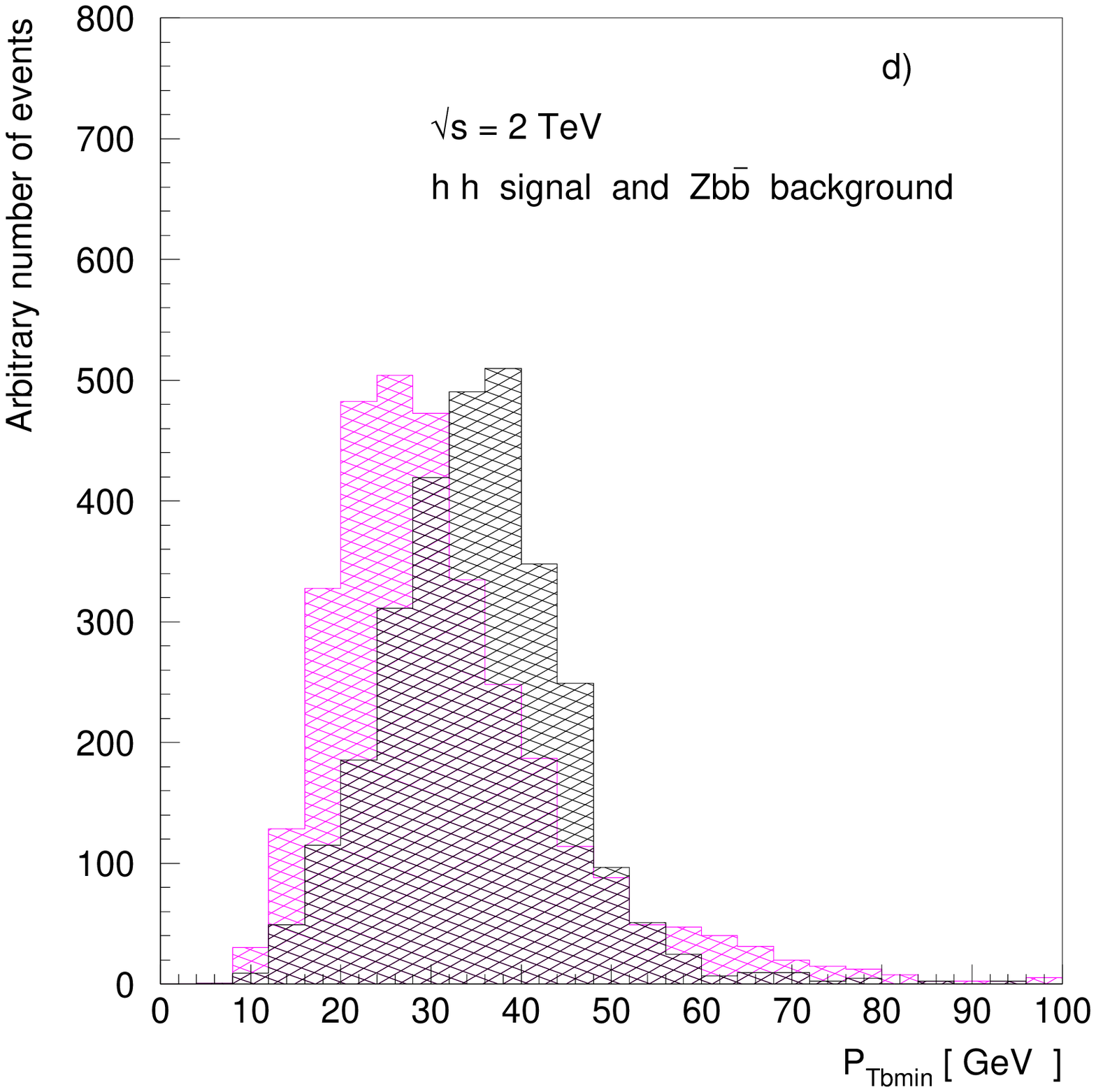 ,width=0.45\textwidth}
      \epsfig{file=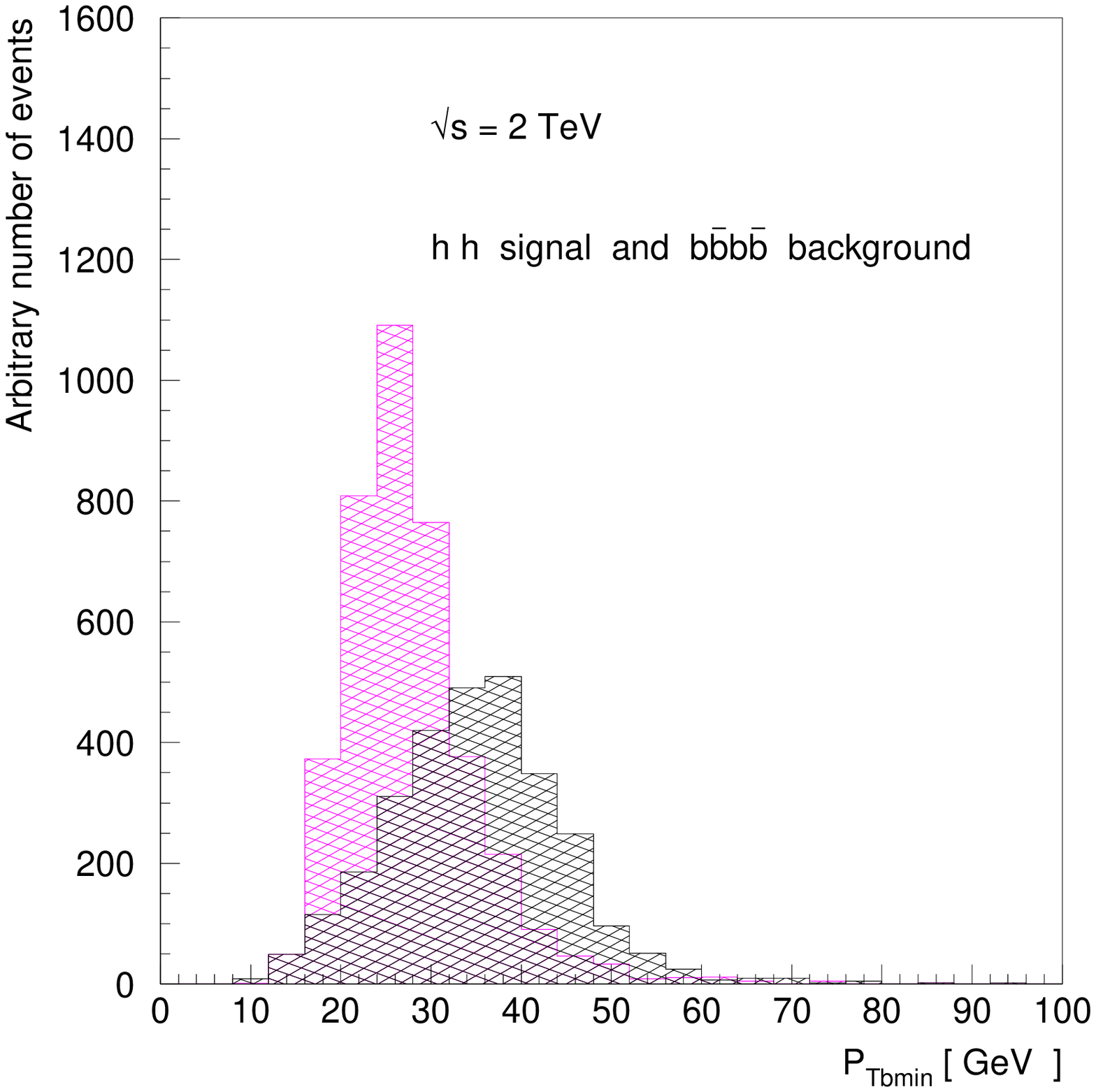,width=0.45\textwidth}}
\vspace*{-0.5cm}
\mbox{\epsfig{file=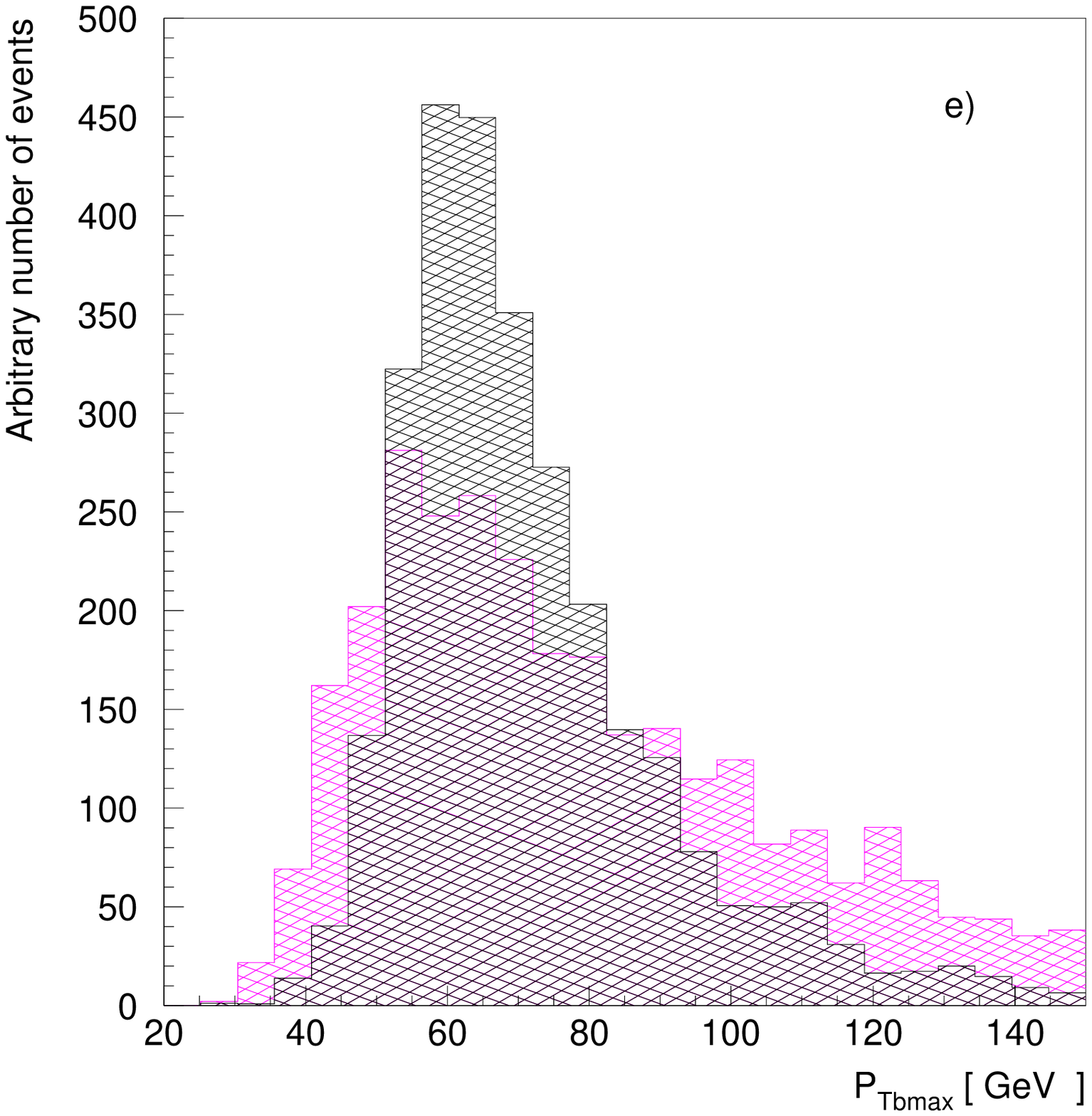 ,width=0.45\textwidth}
      \epsfig{file=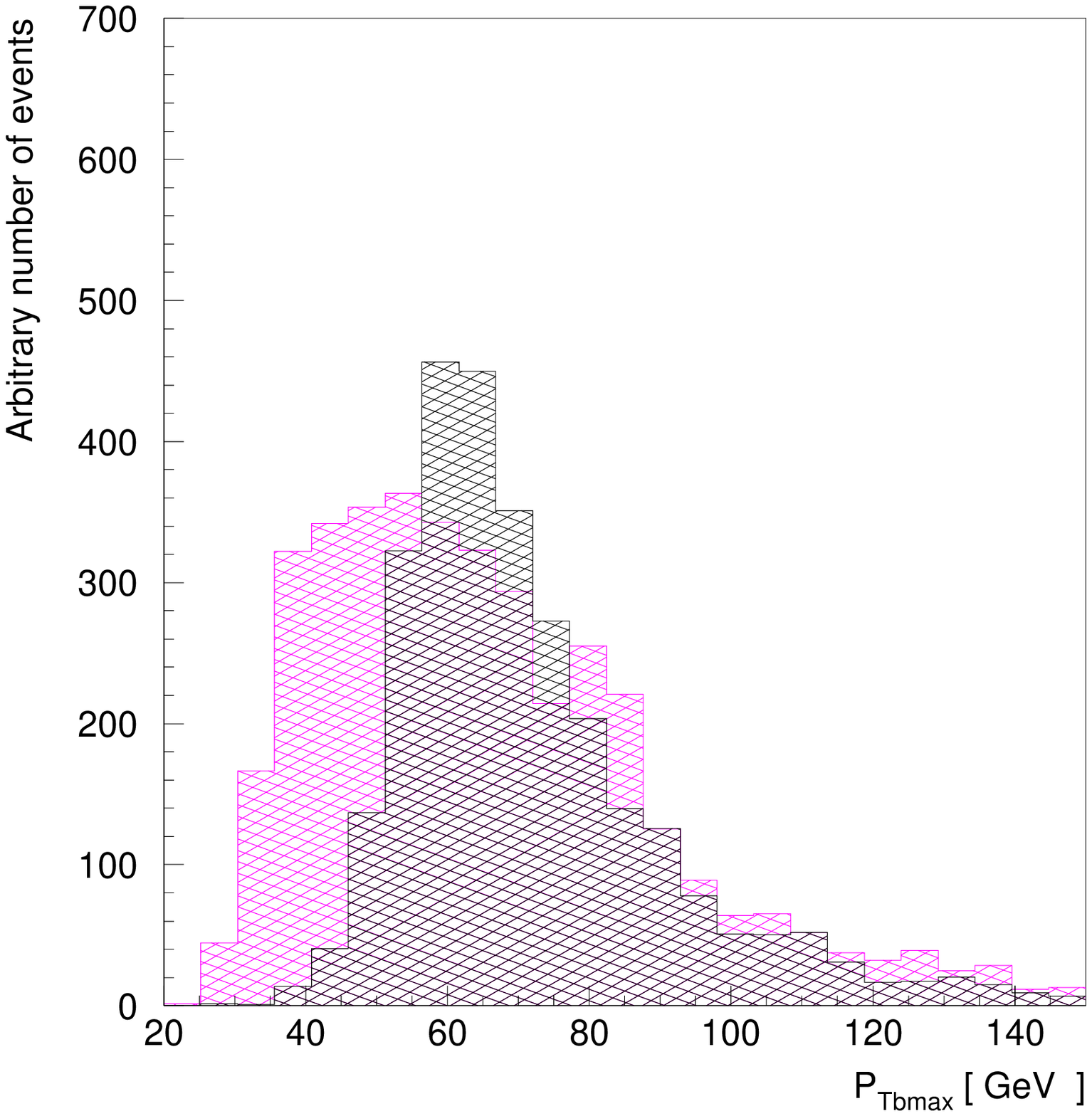,width=0.45\textwidth}}
\vspace*{-0.5cm}
\mbox{\epsfig{file=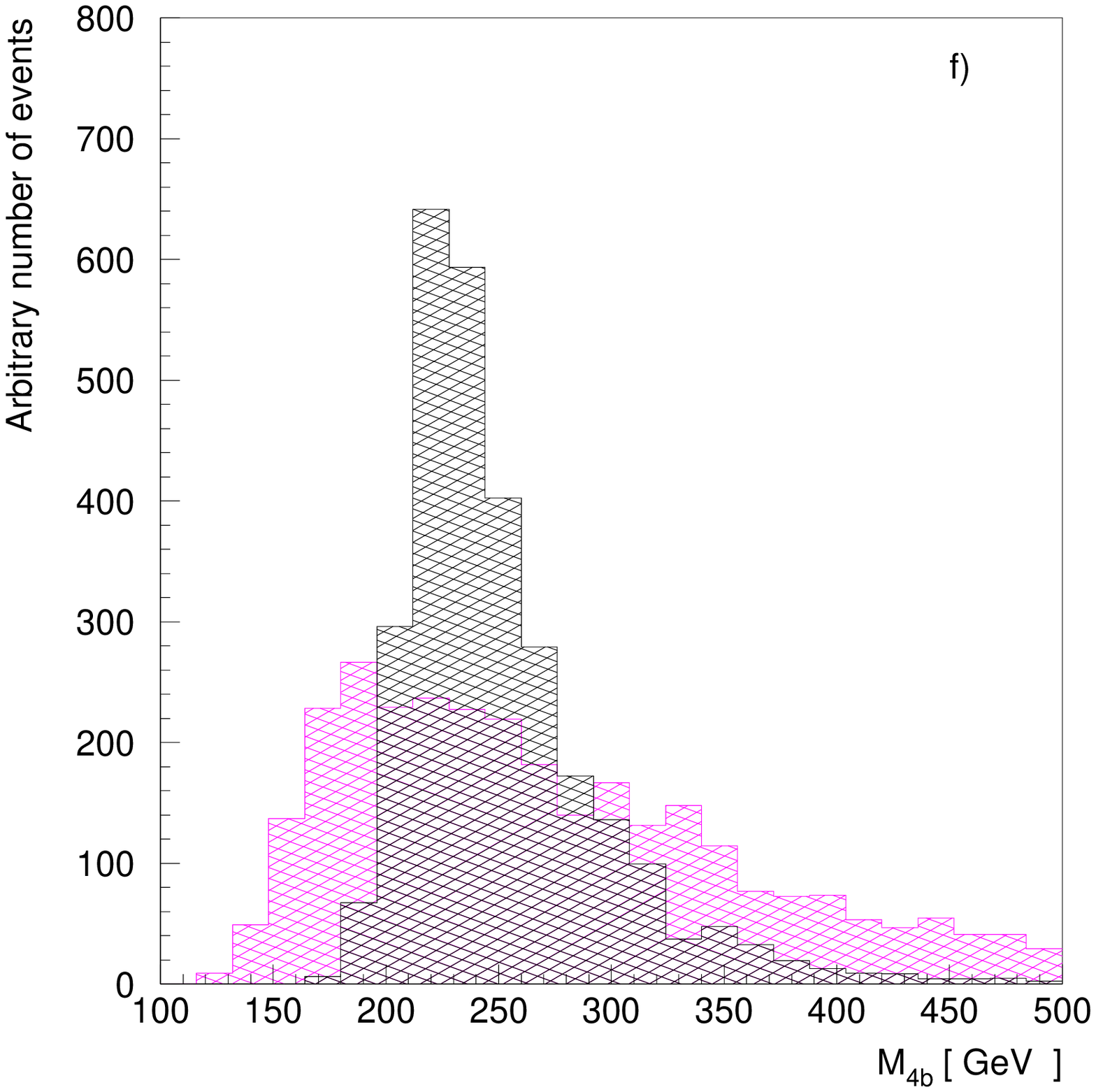 ,width=0.45\textwidth}
      \epsfig{file=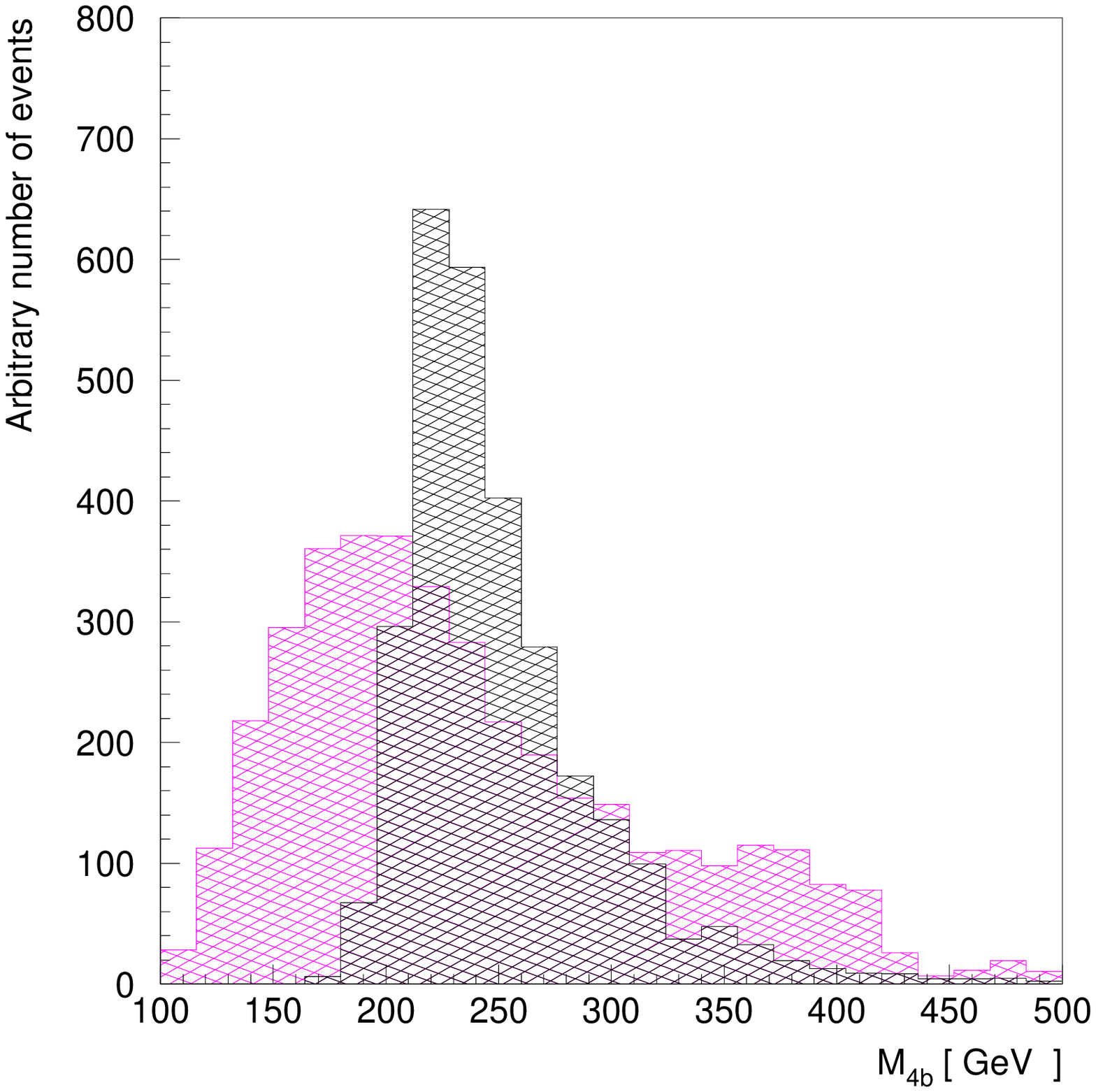,width=0.45\textwidth}}
\protect
\vspace*{0.5cm}
\caption{\label{fig:tev-dist}
Various kinematic distributions of the signal for the
production of a pair of Higgs bosons with mass 120 GeV (dark
histograms and circles), compared to the $Z b \bar{b}$ background
(left column) and the pure QCD $4b$ background (right column). These
results are for the Tevatron; the distributions for the LHC look
similar.}
\vspace*{-0.5cm}
\end{figure}

The jets in the signal are usually quite energetic, with average $p_T$
close to half the mass of the produced Higgs bosons. Since most of
these bosons are produced fairly close to the kinematical threshold,
the two jets in each pair are nearly back--to--back in the transverse
plane. Moreover, since we focus on the production of two Higgs bosons
with (nearly) identical mass, the two $b \bar{b}$ invariant masses
should be equal within errors. These expectations are borne out by our
Monte Carlo simulation. In Fig.~\ref{fig:tev-dist}a--f we show several
kinematical distributions of the signal, compared with the $Z b
\bar{b}$ (left column) and pure QCD $4b$ background (right
column). The signal has been computed for a Higgs boson mass of 120
GeV, assuming negligible contributions from squark loops and a large
value of \tanb\ (so that the signal is mostly due to $b-$quark
loops). If the main contribution to the signal cross section comes
from loops of heavier (s)particles the difference between signal and
background becomes even more pronounced, since the signal is shifted
towards somewhat larger values of $\hat{s}$, and hence has harder
$p_T$ spectra. The results shown in these figures are not normalized,
but include the $b-$tagging efficiency as parameterized in
eq.(\ref{ed0tag}). In detail, we constructed the following kinematical
variables and respective set of cuts for an efficient extraction of
the signal:

\begin{itemize}

\item Reconstructed Higgs boson mass, $M_H$: since we do not attempt
to distinguish $b-$jets from $\bar{b}-$jets, there are three possible
ways to pair up the four $b$ (anti--)quarks in the final state. We
chose the pairing that gives the smallest difference between the
invariant masses of the two pairs. The reconstructed Higgs boson mass
is then defined as $M_H=[M_{b_1b_2}+M_{b_3b_4}]/2$. After resolution
smearing, the distribution in $M_H$ for the signal can be described by
a Gaussian with width $\sigma \simeq \sqrt{M_H}$ (in GeV units),
supplemented by a tail towards small values due to hard out--of--cone
radiation, as well as energy lost in neutrinos produced in the decay
of the $b-$quarks. The same effects also shift the peak of the
Gaussian downwards by about 10\%, as compared to the physical (input)
Higgs boson mass $m_{H,in}$. This is illustrated in
Fig.~\ref{fig:tev-dist}a. We thus define our search window in the
reconstructed Higgs boson mass as 
\be \label{ewindow}
0.9 m_{H,in} - 1.5 \sigma \leq M_H \leq 0.9 m_{H,in} + 1.5 \sigma .
\ee

\item 
The mass difference between the invariant masses of the two pairs
should be small, see Fig.~\ref{fig:tev-dist}b. We thus demand
\be \label{edelmh} 
\Delta M_H =  \left| M_{b_1b_2}-M_{b_3b_4} \right| \leq 2\sigma.
\ee

\item The angles in the transverse plane between the two jets in each
pair should be large: 
\be \label{ephi}
\Delta\phi_{b_1,b_2},\ \Delta\phi_{b_3,b_4} > 1.
\ee
Another feature of the signal is that the two Higgs bosons have
similar $p_T$ (it would be equal in the absence of initial state
radiation), and hence also similar transverse velocities (or
boosts). The two transverse opening angles therefore tend to be
correlated, so that the difference between them is small. We can thus
require: 
\be \label{edelphi}
\left| \Delta \phi_{b_1,b_2}-\Delta \phi_{b_3,b_4}
\right| <1.
\ee
These cuts are illustrated in Fig.~\ref{fig:tev-dist}c.

\item
All four $b-$jets in the signal are fairly hard. We nevertheless found
it advantageous to use slightly different cuts for the softest and hardest
of these jets, with transverse momenta $p_{T,min}$ and $p_{T,max}$, 
respectively, see Figs.~\ref{fig:tev-dist}d,e: 
\ben \label {eptcut} \beq
{\rm Tevatron:} \  p_{T,min} & > M_H/8 +1.25\sigma; \
          p_{T,max} > M_H/8 + 2 \sigma. \\
{\rm LHC:} \ p_{T,min} &> M_H/4; \ p_{T,max}> M_H/4 + 2 \sigma.
\eeq \een
We use harder cuts for the LHC since here it is important to increase
not only the significance but also the signal to background ratio.
For the large number of events expected at the LHC, systematic errors
will play an important role in the observability of the signal;
conclusions based on an analysis of the statistical significance only
may therefore be misleading.

\item 
The $4b$ invariant mass $M_{4b}$ was also found to be important for
two reasons.  First, the signal distribution for this variable is
concentrated around the invariant mass of the Higgs pair, which is
significantly larger than the kinematic minimum implied by the
cuts (\ref{eptcut}) on the transverse momenta of the four $b-$jets; see
Fig.~\ref{fig:tev-dist}f. We therefore require:
\be \label{m4bcut}
M_{4b} > 1.9 M_H - 3 \sigma.
\ee
Moreover, this quantity has been shown to be useful for disentangling
quark and squark loop contributions \cite{bdemn}. Different masses of
the particles in the loop lead to different form factors, and hence to
different dependence on the partonic center--of--mass energy
$\sqrt{\hat{s}} \simeq M_{4b}$.

\end{itemize}

The efficiency of these cuts for various input (search) Higgs boson masses
is listed in the following Tables for the Tevatron and LHC, respectively.
We have used the following sequence of cuts:
\begin{eqalignno*}
\mbox{cut 1:}& \; p_{T,min}, p_{T,max}, \; \mbox{and pseudorapidity}
\; |\eta_b| \leq 2. \\
\mbox{cut 2:}& \; \mbox{cut 1 }+ \Delta M_H. \\
\mbox{cut 3:}& \; \mbox{cut 2 }+ \Delta\phi. \\
\mbox{cut 4:}& \; \mbox{cut 3 + mass window} +M_{4b}.
\end{eqalignno*}
In these Tables we have applied $b-$tagging after all the other cuts.
The signal efficiency refers to the total Higgs pair cross section times
branching ratio for the $4b$ final state. The background efficiency
refers to the background cross section defined through the basic
acceptance cuts ($p_T(b) > 25$ GeV for all four $b$ (anti--)quarks,
and jet separation $\Delta R_{jj} > 0.5$ for all jet pairs).

\begin{table}[htb]
\begin{tabular}{| l | l | l | l | l | l | l | l |}
\hline
\hline
$m_{H,in}$ [GeV] &  & 100 & 120& 140 & 160 & 180  & 200 	\\
\hline
\hline
$\epsilon_{\rm signal}$ [\%]&   &   51 &   56 &   59 &   62 &   65 &   67 \\
$\epsilon_{Zbb}$ [\%]   & cut 1 &  40.1&  27.5&  18.1&  12.5&   8.9&   6.0\\
$\epsilon_{bbbb}$ [\%]&         &  38.6&  19.4&   8.8&   4.6&   2.3&   1.4\\
\hline
$\epsilon_{\rm signal}$ [\%]&   &   49 &   53 &   55 &   56 &   58 &   57 \\
$\epsilon_{Zbb}$ [\%]   & cut 2 &  25.6&  17.6&  11.7&   7.8&   5.6&   3.8\\
$\epsilon_{bbbb}$ [\%]&         &  26.3&  13.1&  5.86&  3.08&  1.52&   .92\\
\hline
$\epsilon_{\rm signal}$ [\%]&   &   40 &   44 &   45 &   46 &   48 &   47 \\
$\epsilon_{Zbb}$ [\%]   & cut 3 &  18.3&  12.5&   8.0&   5.0&   3.4&   2.3\\
$\epsilon_{bbbb}$ [\%]&         &  14.1&  7.04&  3.36&  1.77&   .87&   .55\\
\hline
$\epsilon_{\rm signal}$ [\%]&   &   32 &   35 &   36 &   34 &   36 &   36 \\
$\epsilon_{Zbb}$ [\%]   & cut 4 &  9.87&  3.17&  1.94&  1.22&   .67&   .35\\
$\epsilon_{bbbb}$ [\%]&         & 7.41& 2.71&  .761&  .460&  .195&  .088\\
\hline
$\epsilon_{\rm signal}$ [\%]&   &  1.44&  2.10&  2.53&  2.74&  3.17&  3.30\\
$\epsilon_{Zbb}$ [\%] &$4b$ tag & .460  & .187 & .133 & .0935& .0541& .0314\\
$\epsilon_{bbbb}$ [\%]&         & .292  & .137 & .049 & .0318& .0153& .0072\\
\hline
\hline
$bbbb+Zbb$      &  \# events    &  17.5&   8.1&   3.2&   2.1&   1.1&    .5\\
$Zbb$          &  for          &   2.2&    .9&    .6&    .4&    .3&    .1\\
$bbbb$           &  2 fb$^{-1}$  &  15.3&   7.2&   2.6&   1.7&    .8&    .4\\
\hline
signal [fb] $\times Br(4b)$& 95\% c.l. & 280& 153& 93.8& 78.4& 59.0& 45.3\\
signal [fb] $\times Br(4b)$& $5\sigma$& 575& 413& 268& 229& 179& 
148\\
\hline
\hline
$bbbb+Zbb$      &  \# events    & 219& 101&  40.2&  26.3&  13.2&   6.6\\
$Zbb$           &  for          &  27.2&  11.1&   7.9&   5.5&   3.2&   1.9\\
$bbbb$          &  25 fb$^{-1}$ & 192&  89.9&  32.3&  20.8&  10.0&   4.7\\
\hline
signal [fb] $\times Br(4b)$& 95\% c.l. & 80.2& 37.4& 19.6& 14.6&  9.7&  7.1\\
signal [fb] $\times Br(4b)$& $5\sigma$& 165& 76.9& 40.2& 30.0& 25.5& 19.4\\
\hline
\hline
\end{tabular}
\vspace*{5mm}
\caption{\label{tab:eff-tev} Signal and background efficiencies and 
minimal cross sections for a 95\% c.l. exclusion limit on, as well as a $5
\sigma$ discovery of, Higgs boson pair production at the Tevatron, for
several values of the ``input'' Higgs boson mass $m_{H,in}$. See the text
for a detailed description of the cuts.} 
\end{table}

\begin{table}[htb]
\begin{tabular}{| l | l | l | l | l | l | l | l |}
\hline
\hline
$m_{H,in}$ [GeV] &  & 100 & 120& 140 & 160 & 180  & 200 	\\
\hline
\hline
$\epsilon_{\rm signal}$ [\%]&   &   32 &   33 &   35 &   38 &   35 &   38 \\
$\epsilon_{Zbb}$ [\%]   & cut 1 &  25.1&  14.5&   9.0&   5.7&   3.9&   2.8\\
$\epsilon_{bbbb}$ [\%]&         &  19.4&  10.2&   5.4&   3.1&   2.0&   1.1\\
\hline
$\epsilon_{\rm signal}$ [\%]&   &   30 &   30 &   32 &   34 &   30 &   33 \\
$\epsilon_{Zbb}$ [\%]   & cut 2 &  11.7&   6.2&   3.5&   2.1&   1.3&   1.0\\
$\epsilon_{bbbb}$ [\%]&         & 10.2&  5.08&  2.85&  1.51&   .93&   .56\\
\hline
$\epsilon_{\rm signal}$ [\%]&   &   23 &   23 &   26 &   27 &   23 &   26 \\
$\epsilon_{Zbb}$ [\%]   & cut 3 &   7.8&   4.1&   2.2&   1.2&    .7&    .4\\
$\epsilon_{bbbb}$ [\%]&         &  5.47&  2.83&  1.46&   .83&   .54&   .30\\
\hline
$\epsilon_{\rm signal}$ [\%]&   &   19 &   19 &   20 &   21 &   18.&   19 \\
$\epsilon_{Zbb}$ [\%]   & cut 4 &  4.23&  1.71&   .70&   .56&   .14&   .10\\
$\epsilon_{bbbb}$ [\%]&         & 2.30&  .965&  .508&  .271&  .288&  .119\\
\hline
$\epsilon_{\rm signal}$ [\%]&   &   .16&   .34&   .60&   .90&  1.02&  1.38\\
$\epsilon_{Zbb}$ [\%]  &$4b$ tag& .0334& .0263& .0187& .0190& .0054& .0081\\
$\epsilon_{bbbb}$ [\%]&         & .0139& .0142& .0127& .0112& .0107& .0071\\
\hline
\hline
$bbbb+Zbb$ & \# events    & 4875& 4900& 4337& 3863 & 3501 & 2419\\
$Zbb$      & for          & 305 & 240 & 171 & 174 &  49.0 &  73.6\\
$bbbb$     & 100 fb$^{-1}$&4570 &4660 &4166 &3689 &3452 &2345\\
\hline
signal [fb] $\times Br(4b)$& 95\% c.l. &1212& 570& 290& 171& 118& 
70.1\\
signal [fb] $\times Br(4b)$& $5\sigma$ &3030&1426& 726& 427& 296& 
175\\
\hline
\hline
\end{tabular}
\vspace*{5mm}
\caption{\label{tab:eff-lhc} Signal and background efficiencies and 
minimal cross sections for a 95\% c.l. exclusion limit on, as well as a $5
\sigma$ discovery of, Higgs boson pair production at the LHC, for
several values of the ``input'' Higgs boson mass $m_{H,in}$. See the text
for a detailed description of the cuts.} 
\end{table}

We note that before the $b-$tagging efficiency is included, the signal
efficiency is essentially independent of the Higgs boson mass. At this
stage the signal efficiency is nearly two times higher at the Tevatron
than at the LHC, due to the stronger $p_T$ cuts applied at the
latter. Again before $b-$tagging, the efficiency of the $Z b \bar{b}$
and especially of the pure QCD $4b$ background decreases very quickly
with increasing ``input'' Higgs boson mass, due to the rather soft
$p_T$ spectra of these backgrounds.  However, after the
$p_T-$dependent efficiency for tagging all four $b$ (anti--)quarks has
been factored in, the final signal efficiency increases significantly
with increasing Higgs boson mass, whereas the efficiency for the
dominant (pure QCD) background at the LHC becomes almost independent
of $m_{H,in}$ as long as $m_{H,in} \lsim 150$ GeV. Note also that the
final signal efficiency at the LHC is a factor 3 to 9 smaller than at the
Tevatron, largely due to the worse $b-$tagging efficiency at moderate
values of $p_T$ [cf. eqs.(\ref{ed0tag}) and (\ref{ecmstag})].

These Tables also contain results for the minimal total signal cross
section times branching ratio needed to exclude Higgs boson pair
production at the 95\% c.l., as well as the minimal total cross
section times branching ratio required to claim a $5 \sigma$ discovery
of Higgs boson pair production in the $4b$ final state. We give these
critical cross sections for two values of the integrated luminosity at
the Tevatron, characteristic for the upcoming Run II and for the final
luminosity at the end of the ``TeV33'' run, respectively. In case of
the LHC, we give results for an integrated luminosity of 100 fb$^{-1}$,
corresponding to one year of data at full luminosity.

As already mentioned, systematic uncertainties are a concern
especially at the LHC, where the large signal rate can lead to a very
small signal to background ratio if the significance is defined using
statistical errors only. It can be argued that systematic
uncertainties should be small, since the signal should manifest itself
as a peak in the $M_H$ distribution on top of a smooth background, the
size of which can be determined from side bins. However, given the
rather large width of the peak, the number of available side bins will
in practice be rather limited. We therefore assign an ad--hoc
systematic uncertainty of 2\% on the background estimate, as obtained
by extrapolation from the side bins. We thus require a minimal signal
to background ratio of 0.04 for the 95\% c.l. exclusion limit, and
0.1 for the $5 \sigma$ discovery cross section. This requirement in
fact fixes the critical cross sections at the LHC for $m_{H,in} \leq
180$ GeV. 

The signal efficiencies in these Tables have been computed under the
assumption that the signal mostly comes from $b-$quark loops. Loops of
heavier (s)particles lead to harder $\hat{s}$ distributions
\cite{bdemn}, and thus to harder $p_T$ spectra of the Higgs
bosons. This has several effects. The average transverse momentum of
the $b-$jets will increase, increasing the tagging efficiency as well
as the efficiency of the $p_{T,max}$ cut. On the other hand, the
spectrum of the softest $b-$jet actually becomes softer, since the
probability of substantial cancelations between the $b-$momentum in
the Higgs rest frame and the boost into the lab frame increases with
increasing momentum of the Higgs bosons. Moreover, the average opening
angle between the $b-$jets within each pair becomes smaller, reducing
the efficiency of the angular cuts (\ref{ephi}) and
(\ref{edelphi}). For the given set of cuts the net effect is to
decrease the signal efficiency at the Tevatron by at most 25\%, while
it remains more or less unchanged at the LHC. However, the correct
procedure would be to re--optimize the cuts for the case that the
signal is dominated by loops of heavy (s)particles (top quarks or
$\tilde{t}_1$ or $\tilde{b}_1$ squarks). For example, one should choose
even more asymmetric cuts on $p_{T,min}$ and $p_{T,max}$, such that
the background efficiency remains essentially the same. The final
signal efficiency should then be at least as high as that shown in the
Tables. Given the uncertainties of our calculation, we simply assume
that the efficiencies, and hence the minimal cross sections for 95\%
exclusion and $5\sigma$ discovery, are as listed in the Tables,
independent of the mass of the (s)particle giving the dominant
contribution to the signal.

Finally, it should be noted that we did not include trigger
efficiencies in our calculation. This does not pose a problem for
Tevatron experiments, where a simple 4--jet trigger should be
sufficient, given our $p_T-$cuts (\ref{eptcut}a). However, LHC
experiments will need far higher $p_T$ trigger thresholds for generic
4--jet events; e.g., ATLAS foresees a threshold of 90 GeV per jet
\cite{atlas}, which would remove most of our signal. On the other
hand, (multi)$-b$ triggers are now being studied by the LHC
experiments. We assume that these will indeed be implemented, and
will allow to trigger on our signal events without significant loss of
efficiency.

\section{Potential of Hadron Colliders for Higgs Pair Search}

\begin{figure}[htb]
\hspace*{-1.5cm}
\mbox{\epsfig{file=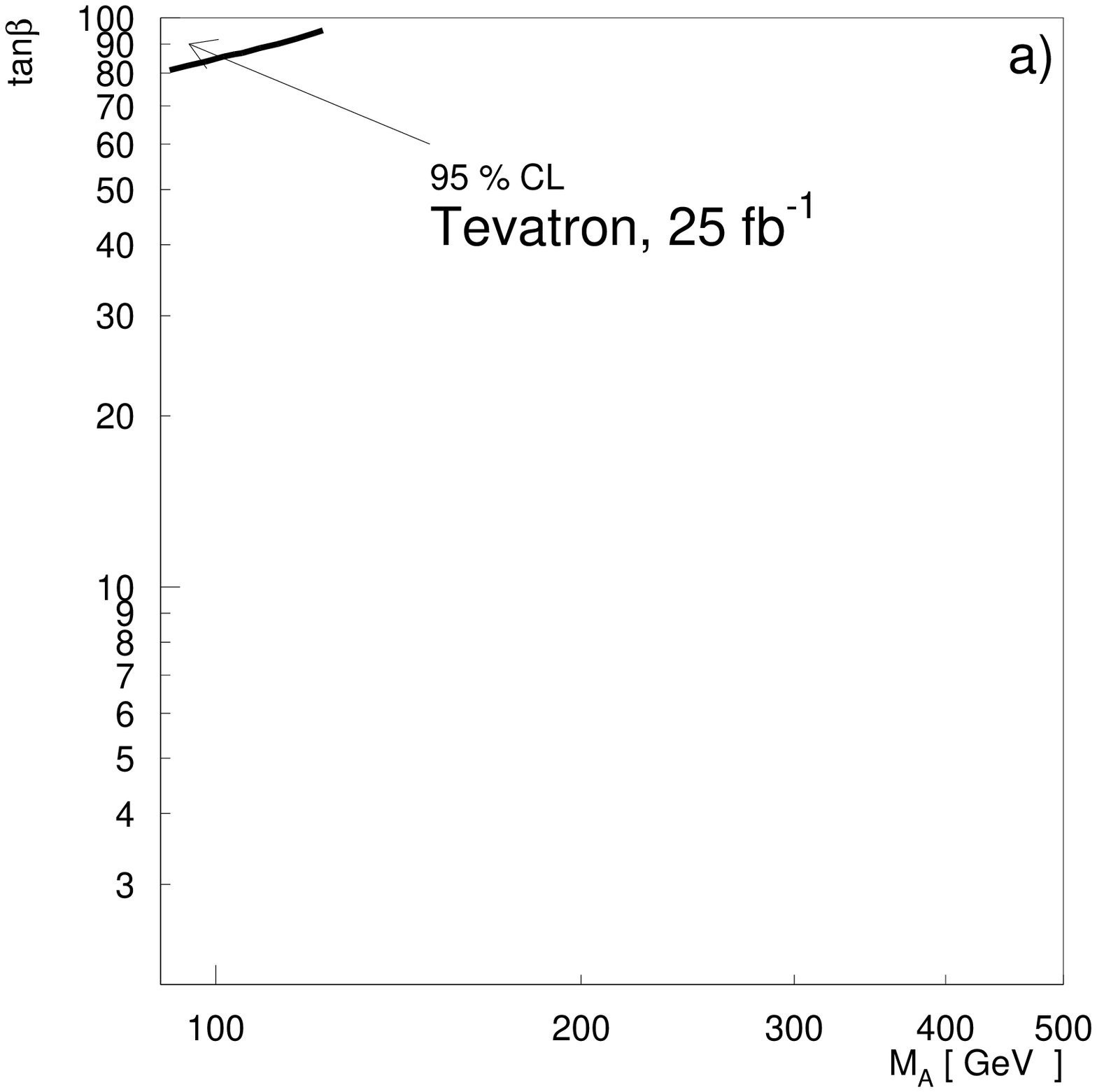,width=0.57\textwidth}
      \epsfig{file=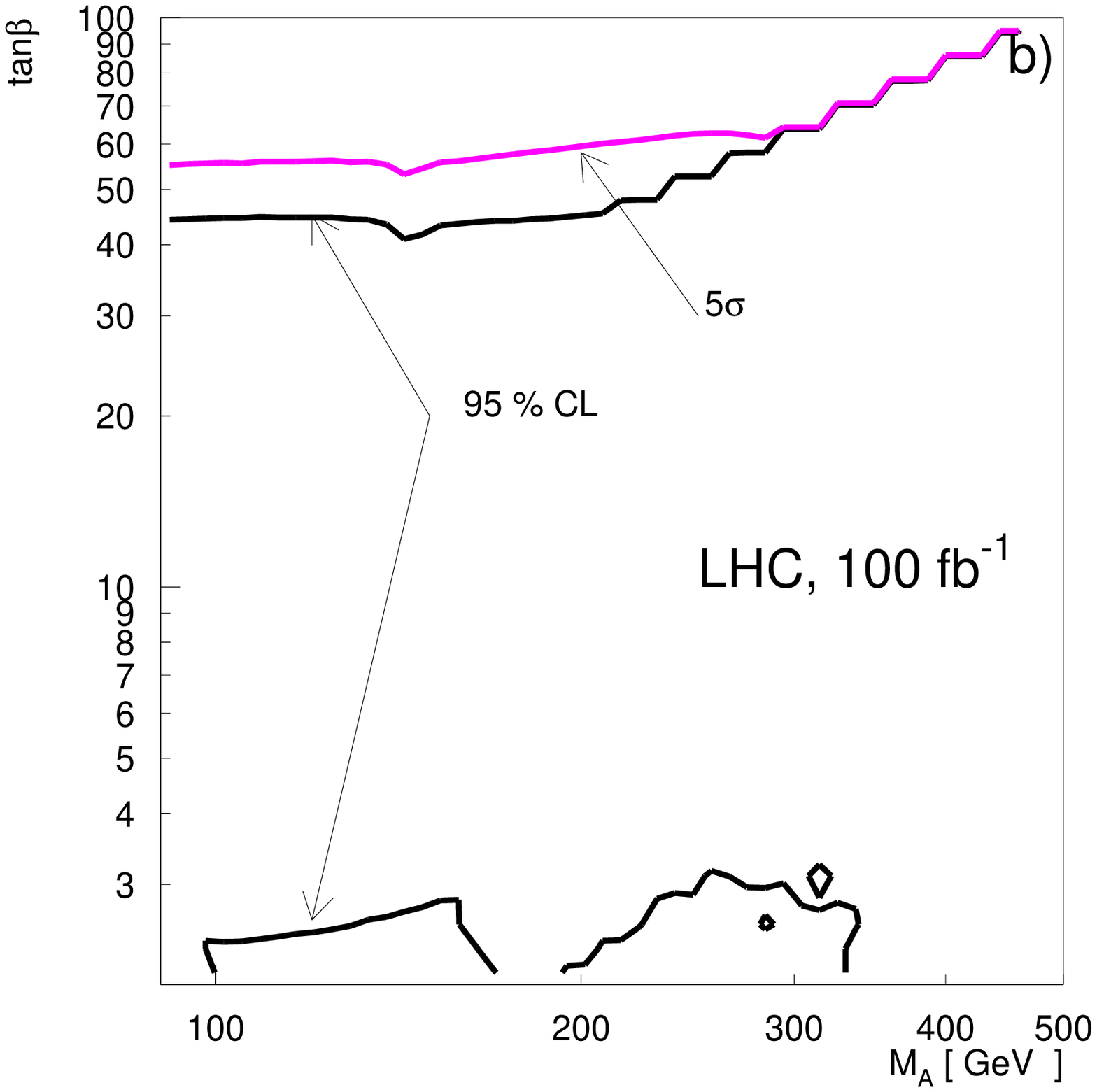,width=0.57\textwidth}}
\caption{\label{fig:nosq-lim}
95\% c.l. exclusion and $5\sigma$ discovery regions for Higgs boson pair
production at the Tevatron  with 25 fb$^{-1}$ of data (a), and 
the LHC with 100 fb$^{-1}$ of data (b), for the case of negligible
squark loop contributions.}
\end{figure}

We are now in a position to estimate the Higgs boson pair production
discovery reach of the Tevatron and the LHC. By comparing the results
of Table~\ref{tab:eff-tev} and Fig.~\ref{fig:cslevel}a, it becomes
clear that in the absence of sizable squark loop contributions to the
signal cross section, the potential of Tevatron experiments for this
search is essentially nil. In contrast, some parts of the $(m_A,
\tanb)$ plane can be covered at the LHC even if squark loop
contributions are negligible. This is illustrated in
Fig.~\ref{fig:nosq-lim}.  Even with 25 fb$^{-1}$ of data, Tevatron
experiments can only exclude a small sliver of parameter space with
$\tanb > 80$, well beyond the upper limit on this quantity in most
implementations of the MSSM ($\tanb \lsim m_t(m_t)/m_b(m_t) \simeq
60$). In contrast, even for this pessimistic assumption of negligible
squark loop contributions, LHC experiments might discover a $5 \sigma$
signal if \tanb\ is large ($\gsim 50$), and can at least exclude some
regions of parameter space where \tanb\ is small ($\lsim 2.5$). Note
that our cuts are not optimized to search for $H \rightarrow hh
\rightarrow b \bar{b} b \bar{b}$ events. By requiring that $M_{4b}$
lies in a narrow window around $m_H$, in addition to cuts similar to
those employed by us, the ATLAS collaboration finds \cite{atlas} a
$\geq 5 \sigma$ signal (with 300 fb$^{-1}$ of data) for $\tanb \leq
3$, for the case $m_H = 300$ GeV.\footnote{Their background
calculation is based on the use of hard $2 \rightarrow 2$ matrix
elements, mostly from $gg \rightarrow gg$, followed by parton
showering. Our calculation, which is based on full $2 \rightarrow 4$
matrix elements, should be more accurate, and might lead to
quantitative changes of the discovery region for $H \rightarrow hh$
events in the $4b$ final state. It should nevertheless be clear that
the presence of a peak in the $M_{4b}$ distribution should make it
easier to look for resonant $hh$ production than to search for the
continuum events which are the main focus of our analysis.}

In ref.\cite{bdemn} it was found that the contribution from squark loops
can exceed that from quark loops by more than two orders of magnitude.
The prospects for Higgs pair searches therefore obviously depend 
crucially on the parameters of the squark sector, in addition to the
parameters $m_A$ and \tanb\ that describe the MSSM Higgs sector at
the tree level. As already stated in the Introduction, squark loop
contributions are large if the relevant lighter squark mass eigenstate
($\tilde{t}_1$ or $\tilde{b}_1$) is light and the trilinear coupling
of this mass eigenstate to Higgs bosons is large. 

In order to illustrate the possible importance of squark loop
contributions, we performed three different Monte Carlo searches of the
three--dimensional parameter space ($m_{\tilde q}, \ A_q, \ \mu$)
describing squark masses and couplings under the assumptions discussed
at the beginning of Sec.~III. In the first and second search we
maximize the square of the coupling of $h$ and $H$, respectively, to
$\tilde{t}_1$ (at small \tanb) or $\tilde{b}_1$ (at large \tanb),
divided by some power of the squark mass, and multiplied with the
branching ratio of the Higgs boson in question into $b \bar{b}$; the
quantity to be maximized is designed to closely track the complete
numerical result for the squared squark loop contribution to $hh$ and
$HH$ production, respectively. In the third search, which is relevant
only for rather small values of \tanb, we maximize the squared $H
\tilde{t}_1 \tilde{t}_1$ coupling multiplied with $Br(H \rightarrow
hh) \cdot Br^2(h \rightarrow b \bar{b})$ and divided by some power of
$m_{\tilde{t}_1}$; here the quantity to be maximized approximates the
squark loop contribution to single $H$ production, followed by
on--shell $H \rightarrow hh$ and $h \rightarrow b \bar{b}$ decays.

These searches were performed for 500 combinations of $m_A$ and \tanb,
placed on a grid with logarithmic spacing. For each point on this grid
we computed the total cross sections corresponding to the three sets
of input parameters found by the three Monte Carlo searches, at both
the Tevatron and the LHC. Finally, we chose the set of input
parameters giving the most significant signal for Higgs boson pair
production. The result does not quite represent the most optimal
situation, since the quantities maximized in the Monte Carlo search do
not completely reproduce the behavior of various contributions to the
total cross section; also, possible interference effects are difficult
to include in such a procedure. Unfortunately, the numerical
calculation of the 6 relevant Higgs pair production cross sections for
a single set of input parameters takes more than two CPU minutes on an
ALPHA--station; a Monte Carlo maximization of the full numerically
computed cross section was therefore not possible with the available
resources. However, we believe that our procedure should reproduce the
maximal cross section compatible with the constraints
(\ref{e1n})--(\ref{e4n}) to within a factor of two or so.

\begin{figure}[htb]
\hspace*{-2cm}
\mbox{\epsfig{file=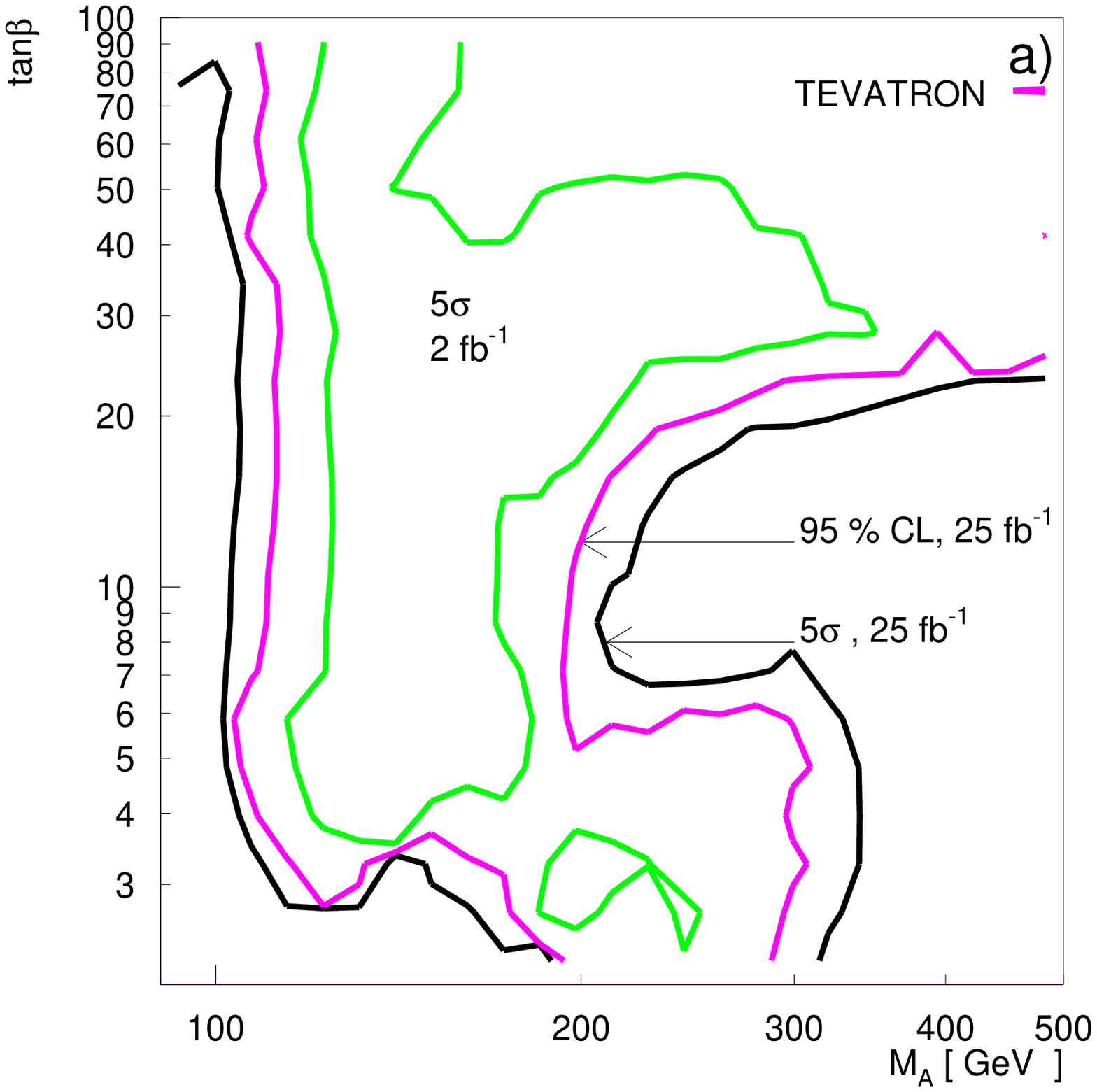,width=0.57\textwidth}
      \epsfig{file=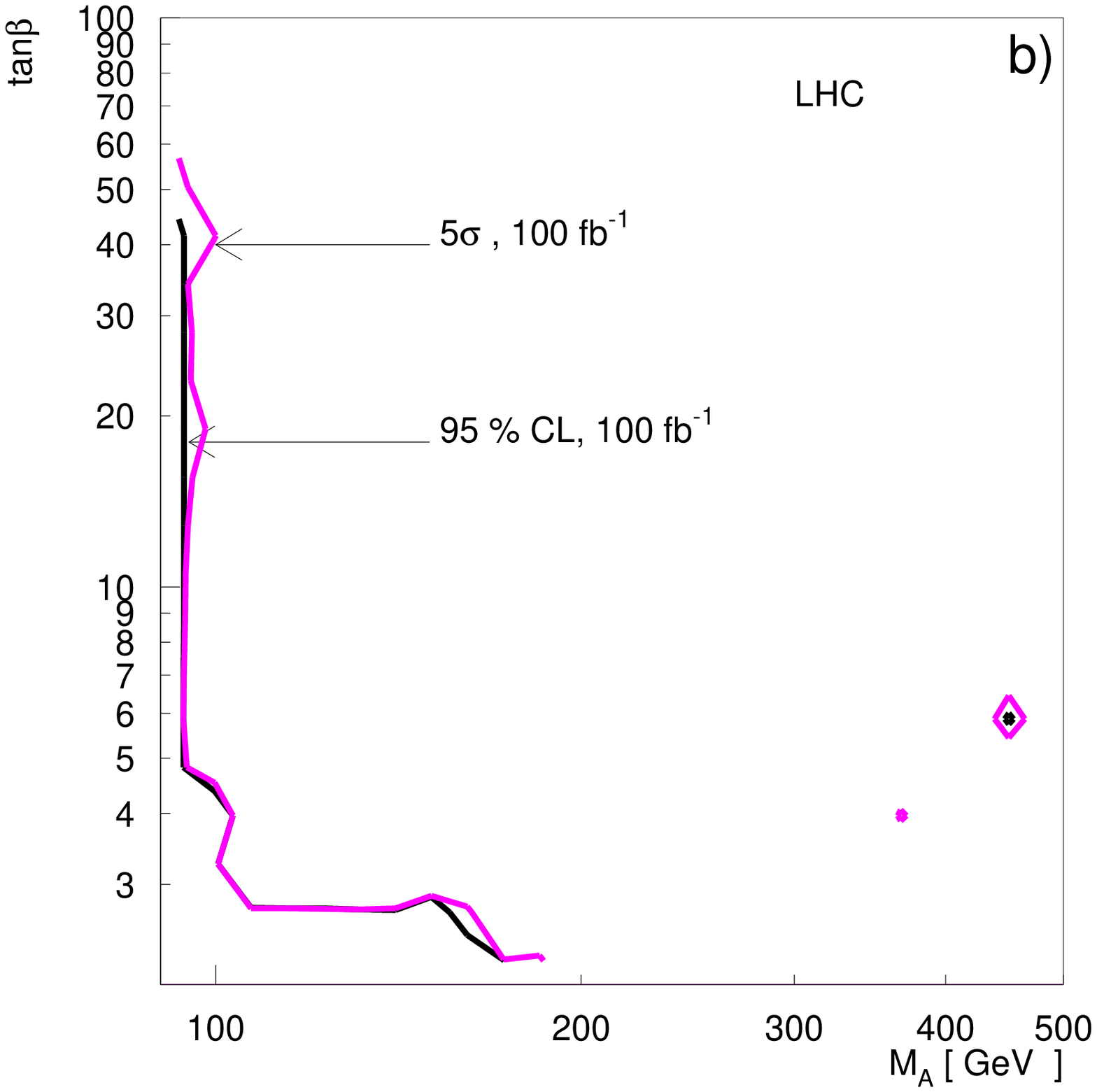,width=0.57\textwidth}}
\caption{\label{fig:sq-lim}
95\% c.l. exclusion and $5\sigma$ discovery regions for Higgs pair
production at the Tevatron (25 fb$^{-1}$) (a) and LHC (100 fb$^{-1}$)
(b), for ``maximized'' squark loop contributions (see the text for
details). The light grey contour in (a) shows the region where a $\geq
5 \sigma$ signal should be detectable at the Tevatron with just 2
fb$^{-1}$ of data. }
\end{figure}

The results are presented in Fig.~\ref{fig:sq-lim}, which shows the
regions that can be probed with 2 and 25 fb$^{-1}$ of data at the
Tevatron (a), and with 100 fb$^{-1}$ of data at the LHC (b). We see that
now virtually the entire part of the $(m_A,\tanb)$ plane still allowed by
the LEP constraints (\ref{e1n}) will give a $ \geq 5 \sigma$ signal at the LHC.
Moreover, the entire region $m_A \leq 200$ GeV, and most of the region
with $m_A \leq 300$ GeV, can be probed at the Tevatron with 25 fb$^{-1}$
of data. Perhaps, the most surprising, and encouraging, result is that a
substantial region of parameter space will give a $\geq 5 \sigma$ signal
at the Tevatron already with 2 fb$^{-1}$ of data! 
This occurs e.g. 
for $m_A = 160$ GeV, $\tanb=10, m_{\tilde q} = 410$ GeV, $A_q=1.15$ TeV
and $\mu = 1$ TeV, leading to a 
$combined \;(hh,HH,AA,hH,hA,HA) \rightarrow 4b$ cross section of
3.3 pb. Another example occurs for $m_A = 170$ GeV, $\tanb=29,
m_{\tilde q} = 326$ GeV, $A_q = 0.78$ TeV and $\mu = 0.98$ TeV, which
gives a signal cross section (before cuts) of 4.7 pb.
This is the first time
that such a robust signal for Higgs boson production at the next run of
the Tevatron collider has been suggested.

\begin{figure}[htb]
\hspace*{-1.5cm}
\mbox{\epsfig{file=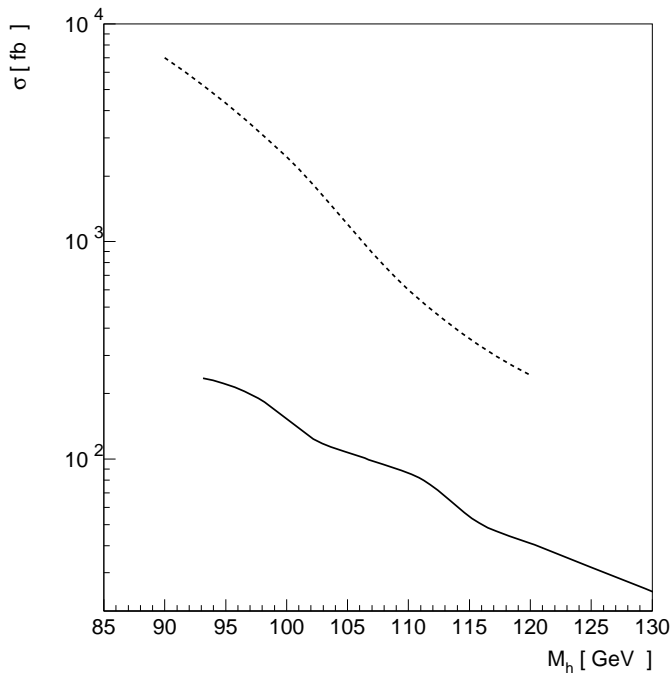,width=0.57\textwidth}}
\caption{\label{fig:cs-mh} Maximal value of the total cross section for
a given value of $m_h$ for the case of heavy squarks (lower curve),
and for the case of ``maximal'' enhancement due to squarks (upper
curve).}
\end{figure}

Our search strategy attempts to maximize the signal for Higgs boson
pair production for given values of $m_A$ and \tanb. This is not the
same as maximizing this signal for fixed mass $m_h$ of the light
CP--even scalar Higgs boson. Nevertheless our search of the squark
parameter space also yielded significantly larger maximal signal cross
sections for fixed $m_h$, as compared to the case of negligible squark
loop contribution depicted in Fig.~\ref{fig:cslevel}. This is
illustrated in Fig.~\ref{fig:cs-mh}, which shows the maximal signal
cross section (before cuts) at the LHC found by our scans of parameter
space, as a function of $m_h$. Attempts to maximize the squark loop
contribution to the signal typically yield $m_{\tilde q} \lsim 400$
GeV; moreover, the choices of $A_q$ and $\mu$ are different from that
that maximize $m_h$. Scenarios with large squark loop contribution
therefore have significantly smaller $m_h$ for given $m_A$ and
\tanb. This explains why the upper curve in Fig.~\ref{fig:cs-mh}
already terminates at $m_h = 120$ GeV, while the lower curve, which is
for the same parameters as Fig.~\ref{fig:cslevel}, extends to 130
GeV. This also explains why the enhancement due to squark loops
becomes smaller as $m_h$ increases.  We should emphasize that cases
with significantly larger enhancement of the signal due to squark
loops may well exist, since, as already noted, our search strategy was
not optimized for maximizing this enhancement for a fixed value of
$m_h$. Nevertheless the qualitative behavior of the true maximum of
the signal should be similar to that depicted in Fig.~\ref{fig:cs-mh}.

As already noted, the results shown in Figs.~\ref{fig:sq-lim} and
\ref{fig:cs-mh} show the most significant signal channel. We found
that, unlike for the case of negligible squark loop contribution, the
most significant signal now always comes from $hh$ production, in some
cases augmented by the production of nearly degenerate Higgs bosons
($hA$ and $AA$ production); however, these auxiliary modes contribute
much less to the total signal, since squark loop contributions to
these modes are absent (for the $hA$ channel) or relatively small (for
$AA$ production).  It is then interesting to investigate the question
whether the production of heavier Higgs bosons, in particular $HH$
production (augmented again by nearly degenerate modes), might also
yield a (less significant, but still) observable signal. 

\begin{figure}[htb]
\hspace*{-2cm}
\mbox{\epsfig{file=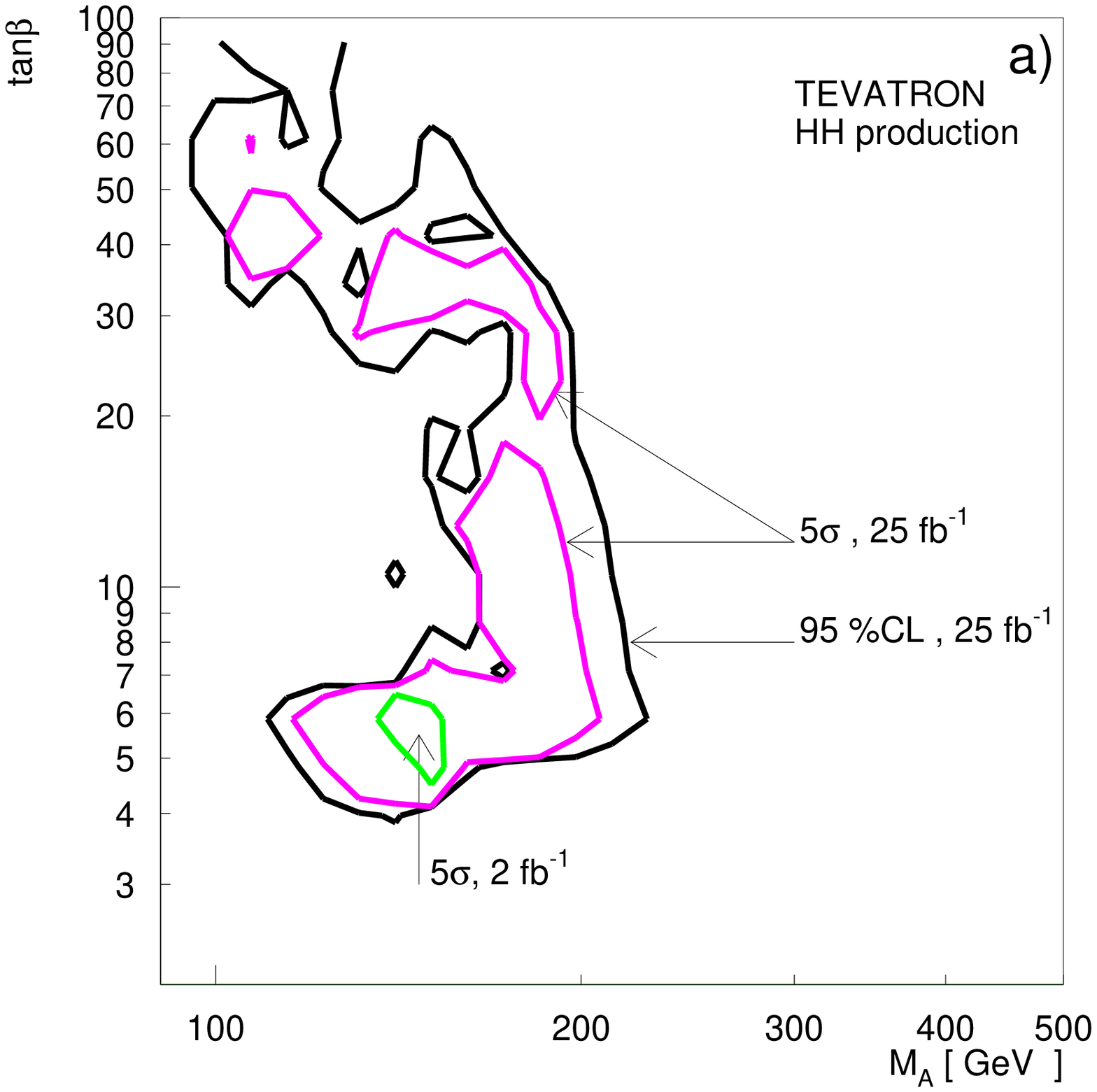,width=0.57\textwidth}
      \epsfig{file=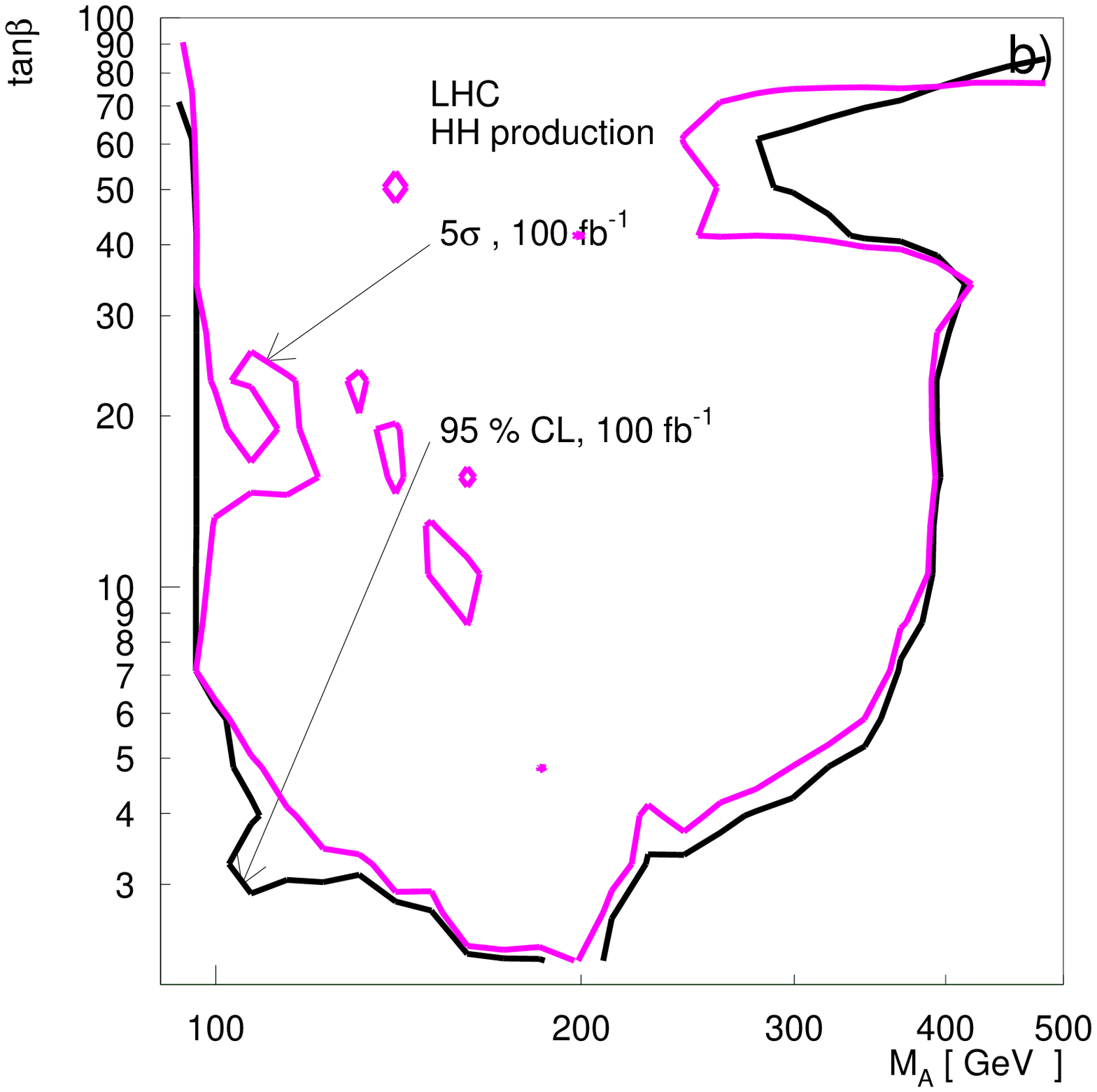,width=0.57\textwidth}}
\caption{\label{fig:sq-lim-hh}
95\% c.l. exclusion and $5\sigma$ discovery regions for heavy Higgs pair
production at the Tevatron (25$fb^{-1}$) (a) and LHC (100$fb^{-1}$)
(b), for ``maximized'' squark loop contributions (see the text for
details). The light grey contour in (a) shows the region where a $\geq
5 \sigma$ signal should be detectable at the Tevatron with just 2
fb$^{-1}$ of data. }
\end{figure}

The answer is shown in Fig.~\ref{fig:sq-lim-hh}, which shows the
potential of the Tevatron and the LHC for observing a signal from $HH$
(heavy Higgs boson) pair production in the $4b$ final state for the
case of ``maximal'' squark loop contribution; as usual, contributions
from nearly degenerate final states are included. We see that a
significant part of the presently still allowed parameter space with
$m_A \lsim 200$ GeV can be probed with 25 fb$^{-1}$ of data at the
Tevatron. LHC experiment will be able to extend this reach over most
of the region where $m_H \leq 2 m_t$, if squark loop contributions are
indeed very large. However, once $m_H > 2 m_t$, the branching ratio of
$H$ into $b \bar{b}$, and hence the signal in the $4b$ final state,
drops quickly. In principle one might then be able to extract a signal
in the $4t$ final state; however, since the decay of each top quark
produces up to three separate jets, the kinematical reconstruction
will be very difficult in this case. Similarly, for $m_A > 200$ GeV,
$m_H \leq 2 m_t$ and small \tanb, $H \rightarrow hh$ decays dominate;
$HH$ production would then lead to final states with up to 8
$b-$quarks! In this case it might be sufficient to simply require
$\geq 5 \ b-$tags in the final state; kinematic reconstruction may not
be necessary to suppress the background.

It should be noted that the squark mass parameters used in
Fig.\ref{fig:sq-lim-hh} usually differ from those used in
Fig.\ref{fig:sq-lim}. Parts of the $(m_A,\tanb)$ plane that appear
accessible in both the $hh$ and $HH$ final states may therefore in fact
not yield simultaneous signals in both channels. However, cases do
exist where two distinct signals can be found, in particular at the
LHC.

\section{Summary and Conclusions}

In this paper we have studied the potential of the Tevatron collider
and the LHC for the search for the pair production of the neutral
Higgs bosons of the MSSM in the $b \bar{b} b \bar{b}$ final state. To
that end, we wrote Monte Carlo event generators for all signal and
(irreducible) background processes. We included initial and final state
showering, hadronization, and heavy hadron decay through an interface
with PYTHIA/JETSET. We also introduced resolution smearing for jet
energies. 

We used these event generators to perform a detailed kinematical
analysis with the aim of extracting the signal, working out the
optimal set of kinematical cuts. The main outcome of this analysis are
values of the minimal total signal cross section times branching ratio
required for a 5$\sigma$ observation of the signal, as well as for
placing 95\% c.l. exclusion limits, at both the Tevatron and the LHC.

In the latter case we found that systematic errors play a crucial
role. Due to the large event rate achievable at the LHC, the 
cross section required for a signal with $5 \sigma$ statistical
significance corresponds to a signal to background ratio of just a few
percent, for Higgs masses below 130 GeV. In our analysis we assigned
a 2\% systematic uncertainty (at the $1\sigma$ level) to the background
estimate. Clearly a detailed experimental analysis will be required to
determine how (un)realistic this assumption is. Unfortunately, such an
analysis may only be possible after LHC experimenters have had
the opportunity to analyze real data on multi$-b$ final states.

We should emphasize here that our signal only includes the pair
production of two neutral Higgs bosons through gluon fusion. Other
contributions to the pair production of neutral Higgs bosons exist
\cite{dkmz}, but have far smaller cross sections. In contrast, the
total cross section for associated Higgs $b \bar{b}$ production can be
large at large \tanb\ \cite{yuan}; this also contributes to $4b$ final
states. However, in this case two of the $b-$jets are quite soft, and
therefore have low tagging efficiencies. Most of the remaining
contribution will be removed by our kinematical cuts. We therefore
believe that our calculation includes the by far most important
contribution to the signal.

Similarly, our background calculation only includes irreducible
backgrounds from pure QCD $b \bar{b} b \bar{b}$ production, as well as
from $Z b \bar{b}$ production followed by $Z \rightarrow b \bar{b}$
decay. We showed that ``fake'' backgrounds should be negligible if the
``false positive'' $b-$tag efficiency of light quark and gluon jets is
a few percent or less. We also estimated backgrounds from
multi--parton ($4 \rightarrow 4$) processes to be far smaller than
those from $2 \rightarrow 4$ processes, and argued that the
potentially sizable background from independent $pp$ interactions
during the same bunch crossing at the LHC can be removed by requiring
that all four $b$ jets come from the same primary vertex.  

We did not attempt to estimate backgrounds from the production of
supersymmetric particles. These frequently produce hard $b-$jets in
their cascade decays \cite{baer}, some of which might even come from
the decay of on--shell Higgs bosons. However, the final state will
then also contain additional energetic particles, in particular two
LSP's. If these are stable or long--lived, the events will tend to
have a large transverse momentum imbalance, unlike our signal. If the
LSP's decay, the events will contain several hard photons, leptons
and/or additional non$-b$ jets, and should thus again be easily
distinguishable from the signal we are investigating. We therefore
believe our background estimate to be reliable (with the caveats
required for any leading order QCD calculation).

The main results of our paper are summarized in
Fig.~\ref{fig:nosq-lim} for negligible squark loop contributions, and
Figs.~\ref{fig:sq-lim} and \ref{fig:sq-lim-hh} for the case of very
large squark loop contributions to the signal. The former case is not
quite the most pessimistic one, since mild destructive interference
between quark and squark loop contributions is possible \cite{bdemn}.
However, we did not find any scenario where this reduces the cross
section by more than 20\% or so; this is hardly significant, given
that our matrix elements are only calculated to lowest nontrivial
order in QCD perturbation theory. Similarly, Figs.~\ref{fig:sq-lim}
and \ref{fig:sq-lim-hh} probably do not quite show the most optimistic
scenarios, since we had to use approximate search strategies of the
squark parameter space when maximizing the signal.

In the absence of substantial squark loop contributions, the prospects
for Tevatron experiments appear to be dim. Even with 25 fb$^{-1}$ of
data only a tiny corner of the currently still allowed MSSM parameter
space can be probed. Even LHC experiments can then only probe scenarios
with $m_A \lsim 300$ GeV and either very large or quite small values of
\tanb.

On the other hand, if squark loop contributions are nearly maximal,
and if it is possible to construct an efficient trigger for events
containing 4 $b-$jets with $\langle p_T \rangle \sim 50$ GeV, LHC
experiments should find a signal for $hh$ production for practically
all allowed combinations of $m_A$ and \tanb; $HH$ production
(augmented by nearly degenerate modes) should be visible for most
scenarios with $m_H \leq 2 m_t$. Moreover, with 25 fb$^{-1}$ of data,
Tevatron experiments would be sensitive to most of the region with
$m_A < 300$ GeV; if \tanb\ is large, even scenarios with $m_A > 500$
GeV might be detectable. Our most exciting result is that a
significant region of parameter space with $m_A \lsim 250$ GeV should
be accessible already at the next run of the Tevatron collider, which
is projected to collect 2 fb$^{-1}$ of data. This seems to be the most
robust signal for the production of MSSM Higgs bosons at the Tevatron
that has been suggested so far.

However, these possibly very large squark loop contribution carry a
price. First, it may not be easy to translate non--observation of a
signal into a bound on MSSM parameter space. Not only the total cross
section, but also kinematical distributions depend on the values of
parameters describing both the Higgs sector and third generation
squarks; even our simplified treatment ended up with five free
parameters. Note that the kinematic characteristics of the signal
depend not only on the masses of the Higgs bosons in the final state,
but also on the masses of (s)particles in the dominant loop
contributions to the signal, which determine the $\hat{s}$ dependence
of the partonic cross section. Secondly, the presence of squark loop
contributions of a priori unknown size might also make it difficult to
use $h-$pair production to constrain the Higgs potential by measuring
the $Hhh$ coupling, as suggested in ref.\cite{dkmz}. On the other
hand, the dependence of both the normalization and the shape of
various differential signal cross sections on numerous MSSM parameters
also means that a positive signal could teach us a great deal about
both the Higgs and the squark sector of the theory.

In this article we have only considered the $4b$ signal for Higgs
boson pair production. Some 15 to 20\% of all Higgs boson pairs will
decay into $b \bar{b} \tau^+ \tau^-$ final states. While subdominant,
this mode will probably offer a better signal to background ratio
after basic acceptance cuts. On the other hand, the presence of
several neutrinos in the final state makes kinematic reconstruction
more difficult. A dedicated signal and background analysis in this
channel might nevertheless prove rewarding. Continuing in the direction
of cleaner final states with smaller branching ratios, one might search
for $b \bar{b} \gamma \gamma$ events. Even though the branching ratio
is now only in the $10^{-3}$ range, an ATLAS study \cite{atlas} found
a better discovery reach for $H \rightarrow hh$ in this channel than
in the $4b$ channel. However, their study assumed that squarks are
heavy; in general, squark loop contributions can reduce the branching
ratios of Higgs bosons into two--photon final states \cite{abdel}.
We conclude that there is considerable room for further studies of
the pair production of neutral Higgs bosons at hadron colliders.

\acknowledgments 

We thank Xerxes Tata and Pedro Mercadante for useful discussions, and
Abdel Djouadi for a careful reading of the manuscript. M.D. thanks
the theoretical particle physics group at the University of Hawaii at
Honolulu as well as the school of physics of the Korea Institute of
Advanced Studies, Seoul, for their hospitality. This work was
supported by Funda\c{c}\~ao de Amparo \`a Pesquisa do Estado de S\~ao
Paulo (FAPESP) and by U.S. Department of Energy under the contract
DE-AC03-76SF00515.

\end{document}